%% file: main.tex
\documentclass[sigconf, 10pt]{acmart}
\pdfoutput=1

\usepackage[utf8]{inputenc}
\usepackage{graphicx}

\usepackage{xcolor}

\usepackage{caption}
\usepackage{subcaption}
\usepackage{float}

\usepackage{amsmath}

\usepackage[colorinlistoftodos]{todonotes}
\usepackage{marginnote}
\let\marginpar\marginnote

\def\ifreviewyes{\iftrue}
\def\ifreviewno{\iffalse}

\ifreviewno
    \newcommand{\myrevcolor} {\textcolor{red}}    
    \newcommand{\myrevcolorb} {\textcolor{blue}} 
    
    \newcommand{\myrevno}[1]  {\todo[color=yellow, nolist, noline]{\textbf{\LARGE#1}}}
\else
    \newcommand{\myrevcolor} {\textcolor{black}}
    \newcommand{\myrevcolorb} {\textcolor{black}}    

    \newcommand{\myrevno}[1]  {  } %empty
\fi

\usepackage{multicol}

\usepackage[short,nodayofweek,level,24hr]{datetime}

\usepackage{float}
\usepackage{enumitem}
\setlist{nosep}
\setlist[itemize,0]{label={},leftmargin=\dimexpr 0.0em}
\setlist[enumerate,0]{label={},leftmargin=\dimexpr 0.0em}

\usepackage{titlesec}
\titlespacing*{\section}{0pt}{0.0\baselineskip}{0.00\baselineskip}
\titlespacing*{\subsection}{0pt}{0.0\baselineskip}{0.00\baselineskip}
\titlespacing*{\subsubsection}{0pt}{0.0\baselineskip}{0.00\baselineskip}

% ...
\usepackage{blindtext}
\usepackage{hyperref}

\usepackage{breakurl} % enable line breaks for long URLs
\usepackage{url}      % activate \url work

\usepackage{algorithmic}
\usepackage[linesnumbered,ruled]{algorithm2e}

\usepackage[]{lineno} 

\usepackage{xspace}

\newcommand{\ie}{{\em i.e.}\xspace}
\newcommand{\eg}{{\em e.g.}\xspace}
\newcommand{\cc}[1]{\mbox{\texttt{#1}}}

\newcommand{\swapclean}{{\cc{swap-clean}}\xspace}

\usepackage{array}
\newcolumntype{?}{!{\vrule width 1.5pt}}

\usepackage{circledsteps}
\newcommand\mycircle[2][]{\Circled[fill color=black,inner color=white,#1]{\sffamily{#2}}}

\newcommand\mycircletwo[2][]{\Circled[outer color=black,inner color=black,#1]{\sffamily{#2}}}

\newcommand{\orcidacm}[1]{\href{https://orcid.org/#1}{\includegraphics[width=8pt]{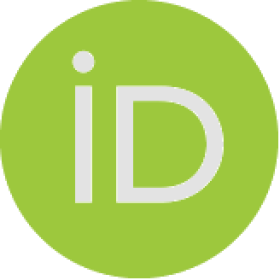}}}

\acmYear{2023}\copyrightyear{2023}
\acmConference [ACM Mobicom '23] {The 29th Annual International Conference on Mobile Computing and Networking}{October 2--6, 2023} {Madrid, Spain}
\acmBooktitle{The 29th Annual International Conference on Mobile Computing and Networking (ACM MobiCom '23), October 2--6, 2023, Madrid, Spain}
\acmPrice{15.00}
\acmDOI{10.1145/3570361.3592518}
\acmISBN{978-1-4503-9990-6/23/10}

\begin{document}
% \begin{NoHyper}
%\vfill
%\raggedbottom
%\flushbottom

\input{section/01title.tex}

\input{section/02author.tex}

\input{section/05abstract.tex}

% \coffeestainD{0.4}{0.5}{90}{3 cm}{1 cm}
\section{Introduction}\label{sec:introduction}
\input{section/10intro.tex}

\section{Background}\label{sec:background}
\input{section/20background.tex}

\section{Observation}\label{sec:observation}
\input{section/30observation.tex}

\section{Mobile-Aware SWAM}\label{sec:design}
\input{section/40design.tex}

\section{Evaluation}\label{sec:evaluation}
\input{section/70evaluation}

\section{Discussion}\label{sec:discussion}
\input{section/75discussion}

\section{Related Work}\label{sec:related}
\input{section/80relatedwork.tex}

\section{Conclusion}\label{sec:conclusion}
\input{section/90conclusion}

%% TODO: When accepted, add this line for Camera-Ready-Version (CRV).
\section*{Acknowledgment}\label{sec:ack}
\input{section/95ack}

\input{section/97reference}

% \end{NoHyper}
\end{document}

%% file: section/01title.tex
\title{SWAM: Revisiting Swap and OOMK for Improving\\ Application Responsiveness on Mobile Devices}
\subtitle{}

%% file: section/02author.tex
\iftrue
\author{%
% Geunsik Lim$^{1 \orcidacm{0000-0003-1845-7132}}$, Donghyun Kang$^{2 \orcidacm{0000-0003-4362-9944}}$, MyungJoo Ham$^{3 \orcidacm{0000-0002-9731-0253}}$, and Young Ik Eom$^{4 \orcidacm{0000-0001-6141-8054}}$
\href{https://orcid.org/0000-0003-1845-7132}{Geunsik Lim}$^{1}$, \href{https://orcid.org/0000-0003-4362-9944}{Donghyun Kang}$^{2}$, \href{https://orcid.org/0000-0002-9731-0253}{MyungJoo Ham}$^{3}$, and \href{https://orcid.org/0000-0001-6141-8054}{Young Ik Eom}$^{4}$
}
\email{{leemgs, yieom}@skku.edu, donghyun@gachon.ac.kr, {geunsik.lim, myungjoo.ham}@samsung.com}
\affiliation{
    \institution{$^{1,4}$Dept. of Electrical and Computer Engineering \quad $^{4}$College of Computing and Informatics}
    \institution{$^{1,3}$Samsung Research \quad $^{2}$Dept. of Computer Engineering}
    \institution{$^{1,4}$Sungkyunkwan University \quad $^{2}$Gachon University \quad $^{1,3}$Samsung Electronics}
    \country{South Korea}
%    \country{ }
}

\fi

%% Case3:  (TEMP)
% \quad 
\iffalse
\author{%
AAAAA AAA$^{1}$, BBBBB BBB$^{2}$, CCCCC CCC$^{3}$, and DDDDD DDD$^{4}$
}
\email{{aaaaa, ddddd}@my1.org, bbbbb@my2.org, {aaaaa, ccccc}@my3.org}
\affiliation{
    \institution{$^{1,4}$Dept. of 1111111111111 \quad $^{4}$Dept. of 222222222222}
    \institution{$^{1,3}$Dept. of 3333333333333 \quad $^{2}$Dept. of 444444444444}
    \institution{$^{1,4}$AAA Institution \quad $^{2}$BBB Institution  \quad $^{1,3}$CCC Institution }
    \country{\#207}
}

\fi

%% file: section/05abstract.tex
\begin{abstract}

%% https://wordcounter.net/
%% ----- Print a date to identify printed papers
% \usepackage[short,nodayofweek,level,24hr]{datetime}
% \colorbox{yellow}{\color{red}Draft, Rev. 131, The build time: \today\ (\currenttime)}

Existing memory reclamation policies on mobile devices may be no longer valid because they have negative effects on the response time of running applications. 
In this paper, we propose SWAM, a new integrated memory management technique that complements the shortcomings of both the swapping and killing mechanism in mobile devices and improves the application responsiveness. 
SWAM consists of (1) Adaptive Swap that performs swapping adaptively into memory or storage device while managing the swap space dynamically, (2) OOM Cleaner that reclaims shared object pages in the swap space to secure available memory and storage space, and (3) EOOM Killer that terminates processes in the worst case while prioritizing the lowest initialization cost applications as victim processes first. 
Experimental results demonstrate that SWAM significantly reduces the number of applications killed by OOMK (6.5x lower), and improves application launch time (36\% faster) and response time (41\% faster), compared to the conventional schemes.
\end{abstract}

% TODO: replace this section with code generated by the tool at https://dl.acm.org/ccs.cfm
\iftrue
\begin{CCSXML}
<ccs2012>
   <concept>
       <concept_id>10003120.10003138.10003141.10010898</concept_id>
       <concept_desc>Human-centered computing~Mobile devices</concept_desc>
       <concept_significance>500</concept_significance>
   </concept>
</ccs2012>
\end{CCSXML}

%\ccsdesc[500]{Human-centered computing~Mobile computing}
%\ccsdesc[500]{Computer systems organization~Embedded systems}
%%\ccsdesc[500]{Computer systems organization~Processors and memory architectures}
%%\ccsdesc[500]{Information systems~Mobile information processing systems}
%\ccsdesc[500]{Software and its engineering~Embedded software}   

%\keywords{memory management, page reclamation, swap, OOMK, application launch time, application response time}

\fi

% addition items: mobile computing, application launching time, application response time
\maketitle

%% file: section/10intro.tex
Today, as the memory capacity in general-purpose computing systems may not scale to memory consumption trends of the contemporary applications, the memory shortage is a well-known factor that negatively affects the overall system performance.
For example, deep learning or cloud services in a server environment require a significant amount of memory to perform their memory-intensive operations simultaneously on large amounts of data, and such applications used to reveal memory shortage problems when their operations require peak memory space. 
Meanwhile, in mobile devices, the memory shortage problem may be more serious because (1) the number of applications that reside in memory increases over time, and (2) the available physical memory space cannot be easily extended to mitigate the problem. 
\begin{figure}
  \centering
  \includegraphics[width=0.99\columnwidth]{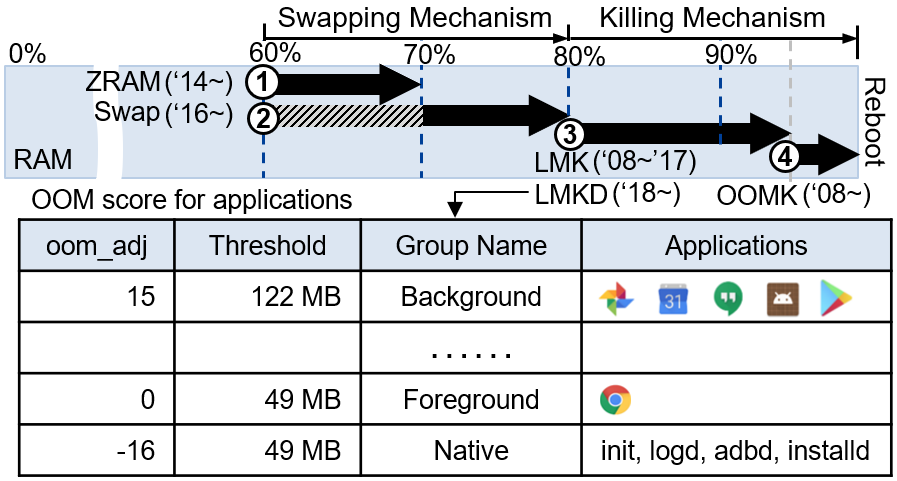}
  % \vspace{0mm}
  \caption{Memory reclamation facilities on mobile devices. OOMK/LMK were enabled since 2008. ZRAM and Swap were enabled since 2014 and 2016, respectively. \myrevcolorb{The striped gray bar indicates that the swap operation is not activated if there is available ZRAM space when both ZRAM and Swap are enabled.} LMK has been replaced by LMKD since 2018. The table refers to the OOM score, memory threshold, and group name of applications that can be killed by LMKD.}
  \label{fig:mem-reclaim}
  % \vspace{0mm} %Put here to reduce
\end{figure}
%\vspace{-1mm} %Put here to reduce
There have been many efforts \cite{atc21-asap, atc18-fasttrack, dac17-smartswap, microcom16-zram-dedup,emsoft16-swap-dram-nvm, atc20-marvin, android-mem-alloc, mobicom15chall, mobicom18android, mobicom15-caredroid, mobisys15-app-rw-isolation} to mitigate this problem in mobile environments.
Figure \ref{fig:mem-reclaim} shows the recent history of memory management technologies in mobile devices. 
As shown in Figure \ref{fig:mem-reclaim}, the operating system on mobile devices has been adopting two types of mechanisms, swapping and killing, since 2008 \cite{opensource-samsung, opensource-huawei, opensource-oppo, opensource-xiaomi, ram-plus, ram-plus-ac, ram-expansion-oppo, android-lmkd, lwn12-us-lmkd, access20-selective-swap-kill, tce13-vnode}.
In mobile devices, when the memory utilization becomes around 60\%, the ZRAM/Swap secures the available memory space by moving some memory pages occupied by processes to the swap area on the memory or storage devices \cite{atc99-swap-compression, microcom16-zram-dedup, atc10-flashvm, tce10-swap-nand, issc13-blockio, dac17-smartswap}. 
However, these approaches negatively affect application response time when the processes re-access the swapped-out pages; it requires extra time to reload the data into memory.
If the memory utilization gets close to 80\% or 90\%, Low Memory Killer Daemon (LMKD) and Out-of-Memory Killer (OOMK) start to kill applications to reclaim a large amount of memory space at \myrevcolorb{once \cite{lwn10-another-oomk, atc16-linux-mm-evolution}}; but, it also suffers from slow launch time when a user relaunches the applications killed by LMKD/OOMK \cite{tecs14-person-opt}.
Note that application response/launch time is very important in that it directly affects the user experience\footnote{User experience refers to the overall experience of perception, reaction, and behavior that users feel and think while using the products and services.} in mobile environments where each device is dedicated to each user.
Therefore, traditional memory management schemes (\eg, Swap and OOMK) should be revisited and reformed to efficiently address this issue.

In this paper, we propose SWAM, which efficiently orchestrates the traditional swapping and killing schemes with three key components. (1) To efficiently handle swapping and eventually to reduce the response time of applications, \emph{\textbf{Adaptive Swap}} dynamically adjusts swap space and provides two independent paths for swapping: the fast path which performs swapping onto memory and the slow path which performs it onto underlying storage devices. 
(2) To secure the available swap space of SWAM, \emph{\textbf{OOM Cleaner}} preferentially reclaims the swapped-out SO pages that include the contents of shared object (.so) files of each application (we call this \swapclean in this paper). 
(3) To reduce application launch time, \emph{\textbf{EOOM Killer}}, which operates based on the mechanism of traditional OOMK, effectively kills the applications with the shorter restart time among the running applications by getting a hint as to their restart time. 

\myrevcolorb{For our evaluation}, we implemented SWAM in an Android mobile device \cite{android-aosp} which occupies 83.8\% (5.3 / 6.3 billion, 2021) of the global market \cite{android-market-share-idc, bankmycell}, and conducted several experiments to confirm the effectiveness of SWAM.
Our evaluation results unequivocally demonstrate that SWAM guarantees available memory space and quick response time for the running applications by gracefully selecting victim applications under several memory pressure scenarios.
Especially, in the experiments on a mobile device with 8 GB RAM \cite{google-pixel6}, SWAM considerably reduces the number of applications killed by OOMK by up to 6.5 times, and improves application launch time (36\% faster) and application response time (41\% faster), compared to the conventional schemes. In summary, the major contributions of this paper are as follows.

\begin{itemize}
\item \textbf{(1)} We first study how the Swap and OOMK of conventional operating systems secure available memory space and analyze the side-effects of their memory space reclamation mechanisms (\S\ref{sec:background} and \S\ref{sec:observation}).
\item \textbf{(2)} We introduce a \myrevcolorb{novel} mobile-aware memory management scheme, SWAM, motivated by our observations, and describe how SWAM complements the shortcomings of the conventional swapping and killing mechanisms in detail (\S\ref{sec:design}).
\item \textbf{(3)} We perform a comprehensive evaluation of SWAM using various mobile workloads on the state-of-the-art mobile devices and compare SWAM with the existing policies including NAND-swap, ZRAM, and ZRAM/NAND-swap (\S\ref{sec:evaluation}).
\end{itemize}

%~\\
%(1) We first study how the Swap and OOMK of conventional operating systems secure available memory space and analyze the side-effects of their memory space reclamation mechanisms (\S\ref{sec:background} and \S\ref{sec:observation}). (2) We introduce a noble mobile-aware memory management scheme, SWAM, motivated by our observations, and describe how SWAM complements the shortcomings of the conventional swapping and killing mechanisms in detail (\S\ref{sec:design}). (3) We perform a comprehensive evaluation of SWAM using various mobile workloads on the state-of-the-art mobile devices and compare SWAM with the existing policies including NAND-swap, ZRAM, and ZRAM/NAND-swap (\S\ref{sec:evaluation}).

% 공간 부족 할경우 이 문구 (생략)
%\myrevcolor{The rest of this paper is organized as follows. \S\ref{sec:background} describes how the operating system secures available memory and storage space at runtime and \S\ref{sec:observation} identifies four significant performance impacts related with memory shortage issues. \S\ref{sec:design} describes the design and implementation of SWAM in detail. \S\ref{sec:evaluation} describes the evaluation results of SWAM and \S\ref{sec:discussion} discusses the general guidelines for using it. \S\ref{sec:related} summarizes related studies, and finally, we conclude the paper in \S\ref{sec:conclusion}.}

%% ###################### ENGLISH ######################

%% file: section/20background.tex
Generally, the Linux kernel reclaims pages (in the order of \emph{anonymous page}, \emph{buffer}, \emph{dentry}, \emph{inode}, and \emph{process}) when the memory space for new allocation is insufficient.
In Linux, the Kernel Swap Daemon (KSWAPD) \cite{dac17-smartswap, emsoft16-swap-dram-nvm} is responsible for reclaiming the memory region that is being used by applications (\ie, pages).
The KSWAPD wakes up periodically, and when the amount of memory space in use exceeds a predefined threshold, it begins memory space reclamation.
In spite of the efforts of KSWAPD, if the available memory space is still insufficient, the kernel inevitably runs OOMK that scans victim processes and kills them to reclaim the whole memory space belonging to them.
Meanwhile, the Android platform adopts and exploits more reclamation steps over the Linux kernel because of the limited hardware resources of mobile devices; each step is triggered in the order of ZRAM, Swap, LMKD, and OOMK (see Figure~\ref{fig:mem-reclaim}).
%ZRAM and LMKD are designed for mobile devices with relatively limited memory space.
%Figure~\ref{fig:mem-reclaim} shows the reclamation steps of Android mobile platform~\cite{android-aosp}; ZRAM, Swap, LMKD, and OOMK execute in sequence to reclaim the memory space.
In this section, we discuss more details on the memory management methods, categorizing them into swapping and killing schemes. 
\subsection{Do Not Kill Processes}
Swap is a virtual memory mechanism that expands available memory space with compressed RAM and/or storage devices.
In this section, we elaborate on RAM-class swap and storage-class swap with their pros and cons.
%~\\
~\\
\textbf{RAM-class swap}: Linux has ZSwap and ZRAM which migrate recently unused pages to a dedicated RAM region \cite{atc16-linux-mm-evolution, linux-zswap, access19-ezswap, linux-zram, tc20-zweilous, rtas20casini, rtas16alhammad}.
ZSwap and ZRAM are the same in that they compress victim pages and store them in a portion of DRAM (\ie, swap area) to reduce the I/O overhead of swapping, while ZSwap also supports swapping onto the storage devices.
%The behavior of ZSwap is different from ZRAM in that ZSwap can move the compressed victim pages to the swap area on storage devices.
%In other words, ZSwap uses DRAM as a write-back cache for the storage devices, but, ZRAM employs only the swap area on DRAM without any swap area in storage devices.

In the early days, the mobile platform adopted ZRAM because of the hardware limitations (\ie, limited lifetime and capacity) of the NAND-based mobile storage media (\eg, eMMC and SD Card); ZRAM employs the swap area only on DRAM without it in the underlying storage devices.
In Android platform, the \textit{ActivityManager}~\cite{android-activity-manager} keeps a list of threads that are unlikely to run but consume more memory space than other threads, so that ZRAM can choose victim pages more efficiently.
ZRAM compresses and transmits the victim pages of the threads identified by the \textit{ActivityManager} into the swap area in memory when the physical memory space is insufficient.
Besides, Android application framework (\eg, \emph{onTrimMemory} \cite{android-dev}) notifies the applications that constantly consume more and more memory space to reduce their memory usage if ZRAM cannot secure the desired free space on time.
The latest ZRAM (Linux 4.14 or later) supports the storage-class swap to use the storage device as a backing-store like ZSwap~\cite{linux-zswap, atc16-linux-mm-evolution}.
%In summary, numerous mobile devices have embraced ZRAM because it is significantly faster than the conventional storage-class swap techniques~\cite{access19-ezswap,tce10-swap-nand}.
%~\\
~\\
\textbf{Storage-class swap:} Various types of systems use storage-class swap because storage media generally costs less than DRAM~\cite{dac17-smartswap, tce10-swap-nand, access19-ezswap}.
Because the swap space in the storage device can be emulated as a memory space, this approach is especially useful in the systems where some processes request a large amount of memory space temporarily~\cite{swapping}.
After a series of innovations on storage technologies, NAND flash is becoming more affordable and higher-performance storage media.
For example, embedded Universal Flash Storage (eUFS) 3.1 (2020) has throughput of 2,100 MB/s read and 1,200 MB/s write, which is 8x and 12x faster, respectively, than the embedded MultiMedia Card (eMMC) 5.0 (2015)~\cite{dram-exchange, eufs-samsung}. 
With higher performance storage devices, mobile platforms have started to employ storage-class swap since 2016~\cite{opensource-samsung, opensource-huawei, opensource-oppo, opensource-xiaomi, ram-plus, ram-plus-ac, ram-expansion-oppo}.
Note that the improvement in the robustness of NAND flash storage devices, assisted by over-provisioning and wear-leveling techniques, has also enabled the storage-class swap to be employed in mobile devices as well.

\begin{figure}
  \centering
  \includegraphics[width=0.99\columnwidth]{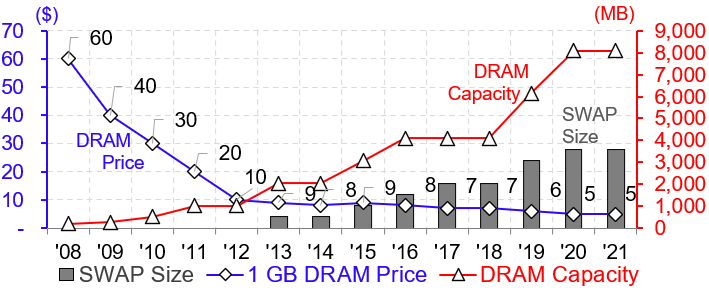}
  %\vspace{0mm} %Put here to reduce
  \caption{The trends in the price of DRAM (1 GB) and the size of DRAM along with the amount of swap space in DRAM on mobile devices}
  \label{fig:ram-swap-change}
\end{figure}
\vspace{1mm} %Put here to reduce

\subsection{Do Kill Processes}
With the mobile platform, the Linux kernel may secure memory space by killing low-priority processes.
In this section, we elaborate on user-mode LMKD and kernel-mode OOMK.

LMKD kills user processes according to the \textit{OOM score} of applications~\cite{android-lmkd, lwn12-us-lmkd} when the available memory space gets smaller than the pre-defined threshold (see Figure \ref{fig:mem-reclaim}).
When LMKD fails to secure enough memory space, the Linux kernel triggers OOMK.
OOMK sequentially kills running processes based on the following heuristics~\cite{atc20-marvin, lwn10-another-oomk}. 
(1) It identifies candidate processes that have a large number of pages, by calculating the amount of memory space allocated to each candidate process.
(2) Then, it excludes the following processes from the candidates: long-running processes in a batch way, processes that fork a few child processes, processes with root permission, processes in accessing hardware resources, and Init/Systemd (PID 1).
(3) Finally, it kills the lowest priority processes one by one from the remaining candidates until enough memory space is secured.

Note that most applications in mobile devices adopt a server-client model and each application runs in a client mode that sends and receives data to/from the remote servers to provide its service \cite{mobicom18android}.
Therefore, killing operations caused by LMKD or OOMK may have tolerable impacts on user experience because each application can restore its state later by connecting to its server.
%% ###################### ENGLISH ######################

%% file: section/30observation.tex
\begin{figure}
\centering
        \begin{subfigure}[b]{0.12\textwidth}
                \centering                \includegraphics[width=\linewidth,height=3.2cm]{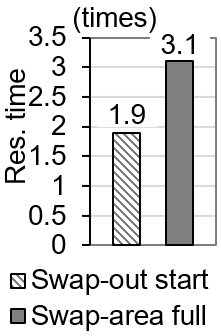}
                \vspace{-4mm} %Put here to reduce                   
                \caption{Swapping}
                \label{fig:observation01}
                \vspace{0mm} %Put here to reduce              
        \end{subfigure}\hfill
        \begin{subfigure}[b]{0.11\textwidth}
                \centering                \includegraphics[width=\linewidth,height=3.2cm]{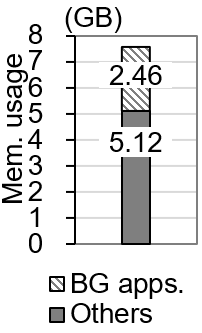}
                \vspace{-4mm} %Put here to reduce                   
                \caption{Memory}
                \label{fig:observation02}
                \vspace{0mm} %Put here to reduce                
        \end{subfigure}\hfill
        \begin{subfigure}[b]{0.12\textwidth}
                \centering                \includegraphics[width=\linewidth,height=3.2cm]{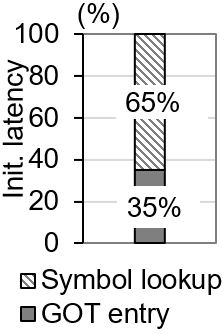}
                \vspace{-4mm} %Put here to reduce                   
                \caption{SO-symbol}
                \label{fig:observation03}
                \vspace{0mm} %Put here to reduce              
        \end{subfigure}\hfill
        \begin{subfigure}[b]{0.12\textwidth}
                \centering
               \includegraphics[width=\linewidth,height=3.2cm]{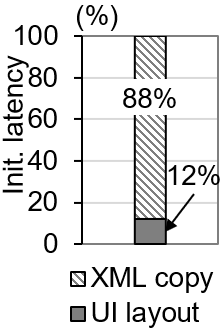}
                \vspace{-4mm} %Put here to reduce                  
                \caption{XML-UI}
                \label{fig:observation04}
                \vspace{0mm} %Put here to reduce    
        \end{subfigure}
        \vspace{0mm} %Put here to reduce
        \caption{\myrevcolorb{The 3 kinds of performance impacts (i.e., response time, memory usage, initialization latency) with 4 sources (e.g., swap, memory, SO, and XML) caused by conventional swapping and killing operations}}
        \label{fig:observation}
        \vspace{0mm} %Put here to reduce
\end{figure}
%~\\
Figure~\ref{fig:ram-swap-change} illustrates the trends of the memory technology in terms of price and capacity in mobile devices; the price is plummeting while the capacity is exploding over time~\cite{dram-exchange, ram-plus, ram-plus-ac, ram-expansion-oppo}.
Unfortunately, today's mobile devices still need the aforementioned memory reclamation operations even though they have enriched DRAM capacity. 
The reason behind this is that the application development paradigm in mobile environments has shifted from small memory consumption toward huge ones; contemporary applications require more and more memory space.
In addition, users usually execute applications (\eg, \emph{video}/\emph{audio players}, \emph{internet}, and \emph{social media}) and use them without terminating until rebooting or running out of battery.
As a result, even with massive amounts of DRAM space, the memory shortage problem still persists.
On the basis of swapping and killing schemes, memory management policies for mobile platforms have extensively been investigated. 
But, unfortunately, they barely focused on the side effects of resource scheduling that may induce negative impacts in mobile systems.

In this section, we introduce \myrevcolorb{three} kinds of major performance impacts with the explanation on the related side effects. To deep dive into the behaviors of traditional swapping and killing operations, we conducted various evaluations while collecting raw data using \emph{dumpstate}, \emph{dumpsys}, and \emph{adb} commands. First, we conducted evaluations on four popular commercial mobile devices (e.g., Samsung Galaxy S9+, S10, S20, and S21 \cite{galaxy-s}) in manufacturers' initial states to understand the overheads caused by the memory shortage problem. Figure~\ref{fig:observation} shows the \myrevcolorb{average performance} impacts measured on the four commercial mobile devices. \myrevcolorb{We} measured the response time of applications (Figure \ref{fig:observation01}) and the total memory usage of applications (Figure \ref{fig:observation02}) when the available memory space is insufficient with the boot-up step. 
Next, we measured the latency derived from the SO-symbol lookup (Figure \ref{fig:observation03}) and XML processing (Figure \ref{fig:observation04}) by running top-ranked representative applications from \cite{google-play}.
\myrevcolorb{We} first targeted 15 top-ranked representative applications \cite{google-play} to assure evaluation reliability, because the evaluation results may be different according to the design structure of SO and the operation pattern of XML-UI. To the best of our knowledge, the cost of GOT entry and Symbol lookup is strongly dependent on the design structure of SO files. Therefore, \myrevcolorb{for SO-symbol lookup experiment (Figure \ref{fig:observation03})}, we additionally selected 45 applications \cite{google-play} to investigate a wider range of applications. In summary, we targeted about 60 applications since the creation cost of GOT entry and Symbol lookup is diverse according to the design structure of SO files, and Figure \ref{fig:observation03} shows our test results on average. Finally, we studied the XML-UI operation patterns based on 15 representative applications that use the UI interfaces intensively (Figure \ref{fig:observation04}).

\subsection{Delays in Response Time}\label{observ:swap-cost}
For fast response time, users of mobile devices commonly tend to keep launched applications in memory until rebooting or running out of battery, even though the applications are not frequently used.
This trend indicates that the memory space can get steadily exhausted in that the number of active applications increases over time. 
If the memory space is insufficient for a new application to start execution, the platform triggers swapping operations to intelligently handle the memory shortage issue.
In this case, the swapping operations negatively influence the response time of the running applications.
Interestingly, the same phenomena can occur right after the boot-up step because the platform starts many applications at once in the background.
To verify our intuition, we investigated how the response time is affected by the swapping operations using \emph{dumpsys} and \emph{adb} commands \cite{android-dev}.

Figure \ref{fig:observation01} illustrates the evaluation results of how the response time of running applications is fluctuated after the first swap-out operation; the response time had been measured after completing the boot sequence and launching applications. As shown in Figure \ref{fig:observation01}, the response time of an application increases by 1.9 times on average when it meets the first swap-out operation.
Meanwhile, if the swap area ends up full due to high memory consumption of applications, the response time increases by 3.1 times; also, it may eventually trigger OOMK operation to secure the available memory space. 

\textbf{Observation 1}: The impacts of swapping on the response time are not negligible and an effective swapping mechanism is more important in modern mobile devices. 

\subsection{Memory Space Consumption}\label{observ:bg-service}
As we know, the Android platform allows running a set of background applications to handle its own functionalities (\eg, \textit{notification, network communication, I/O operation, location information transmission, and data collection}) \cite{android-noti, android-services, mobile-computing-09cont, mobicom18android, mobicom14use, mobicom17-ad-measure}. 
Such background applications can cause a negative impact on the response time of running foreground applications in that they occupy large amount of memory space after booting \cite{mobicom15-caredroid} and may lead to frequent memory reclamation.
To confirm our intuition, we measured how much memory space would be consumed by the background applications using the \emph{dumpstate} command.

Figure \ref{fig:observation02} shows the amount of memory usage of the background applications running on mobile devices. 
As shown in the figure, the background applications, which run without any interaction with users, are allocated about 31\% of the total memory space (2.46 GB of the total 8 GB).
However, such memory space consumption causes two problems because it can frequently trigger operations for memory reclamation.
First, the delay incurred by memory reclamation can exacerbate the launch time and response time when a user runs an application in the foreground.
Note that, as such interferences get more frequent over time, users may feel that their applications are not working in a normal way.
Second, if the memory space occupied by the background applications is reclaimed by OOMK to secure available memory space, it may have a serious impact on the platform service due to the absence of background functionalities.
Therefore, it is important to secure memory space without killing the background applications.

\textbf{Observation 2}: Efficient memory space management is needed not only to support a decent level of user experience but also to keep up with the platform service.

\begin{figure}
  \centering
  \includegraphics[width=0.99\columnwidth]{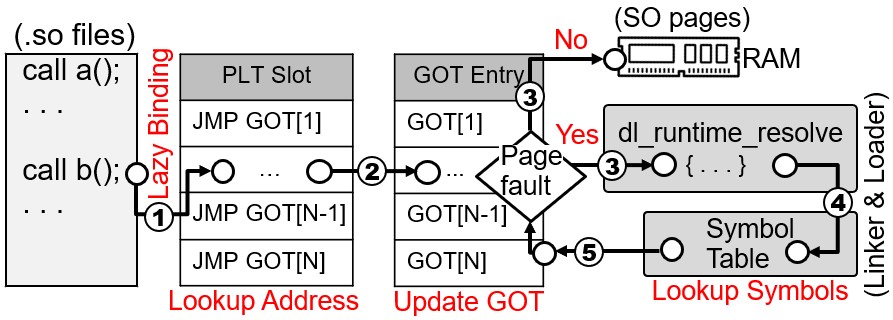}
  \vspace{0mm} %Put here to reduce  
  \caption{Symbol lookup procedure for SO files}
  \label{fig:symbol-lookup}
  \vspace{0mm} %Put here to reduce 
\end{figure}

\subsection{SO-Symbol Lookup Cost}\label{observ:so-lookup}
In Android, native SO (shared objects, .so) files \cite{android-ndk} help to save the memory resources by sharing the memory space of SO files among the applications.
In this section, we first describe how SO files are handled to understand the cost incurred by sharing SO-symbols.

Figure~\ref{fig:symbol-lookup} shows how applications look up symbols of SO files on memory space.
Applications generally do not include symbols of SO files in their own \emph{.text} area so that memory space for such symbols can be shared across applications; thus, applications must use the symbol table to keep all memory addresses of the symbols at run time.
As shown in Figure~\ref{fig:symbol-lookup}, to conserve memory space, the platform employs \textit{lazy binding} policy \cite{linker-loader} that delays the creation of entries in the \emph{Procedure Linkage Table} (PLT) or \emph{Global Offset Table} (GOT) until one of the SO-symbols in the symbol table is accessed.
When a symbol is initially accessed, the \emph{linker} and \emph{loader} \cite{linker-loader} start to scan all the symbols so as to create the entries in PLT and GOT.
To confirm the time consumed by the lookup procedure for SO files, we measured the latency of each step taken during initial access and discovered that symbol lookup (\mycircletwo{\scriptsize{\textbf{3}}}, \mycircletwo{\scriptsize{\textbf{4}}}) and GOT entry generation (\mycircletwo{\scriptsize{\textbf{5}}}) use 65\% and 35\% of the total SO-symbol lookup time, respectively (See Figure ~\ref{fig:observation03}).

Now, let us observe how many SO files are actually shared across applications.
For our observation, we investigated the SO files used in 60 top-ranked mobile applications mentioned in \cite{google-play}.
Surprisingly, we found that, among the 60 applications, only 5 applications (8\%) are sharing some SO files.
In other words, the majority of the applications (92\%) never share their SO-symbol files with others, despite the fact that they incur SO lookup costs when accessing the SO pages which is not loaded in memory space or swapped out into swap space. 
These non-shared SO pages occupied 36\% of the total memory space in our experiment.

\textbf{Observation 3}: Considering that the SO-symbol lookup cost is not negligible while most applications do not share the SO files, it is important which type of SO files to select first as victims for swap-out or memory reclamation.

\subsection{XML Processing Cost}\label{observ:xml-cost}
Android applications generally include Java bytecode and the resource files (\eg, drawable, raw, XML, font, etc.).
The Android Asset Packaging Tool (AAPT) \cite{android-aapt} bundles the resource files into resource packages. At this time, the AAPT performs conversion processes described in XML files (\eg, \emph{animated vector drawable files} and \emph{XML layout attributes}).
When the application framework creates a view hierarchy from the XML file \cite{android-hierarchy-viewer} to load the corresponding application, it uses the compiled XML files on-the-fly. 
Therefore, complex XML files involving the view hierarchy with deep and diverse subsets significantly deteriorate the application load time.
We have observed the XML processing cost by measuring the XML-UI conversion latency at launch time with 15 top ranked applications from \cite{google-play}.
Figure \ref{fig:observation04} presents our evaluation results.
Surprisingly, the application framework consumes approximately 12\% of total XML-UI conversion time on average to interpret the UI layouts of the launched application from the compiled XML files.
Meanwhile, OpenGL ES, for drawing the GUI interface, spends 88\% of the time on average to write the XML data into GPU memory.
\begin{figure}
  \centering
  \includegraphics[width=0.99\columnwidth]{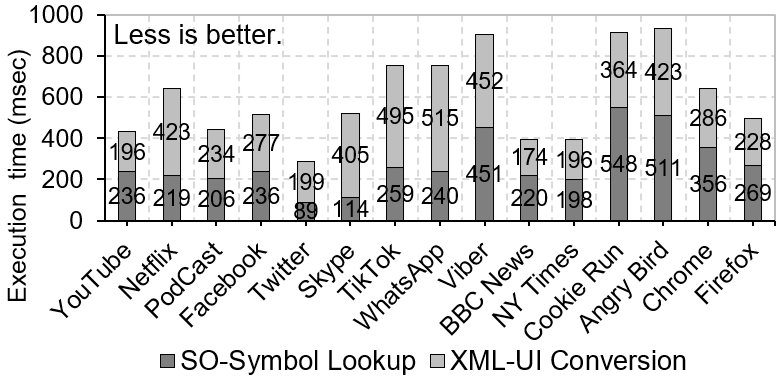}
  \caption{SO-Symbol Lookup and XML processing cost for launching applications in a system with NAND-swap}
  \label{fig:observation-symbol-xml}
\vspace{1mm} %Put here to reduce too much white space after your table
\end{figure}
~\\
\textbf{Observation 4}: The XML processing cost is not negligible anymore in modern mobile devices because the XML files of mobile applications tend to become more complex nowadays.
\subsection{Effect of SO-Symbol Lookup and XML Processing on Application Launch Time}\label{observ:so-xml-cost}
To clearly understand the cost mentioned in Observation 3 (\S\ref{observ:so-lookup}) and 4 (\S\ref{observ:xml-cost}), we measured the launch time of applications, which includes \emph{SO-Symbol Lookup} and \emph{XML-UI conversion} time.
As shown in Figure~\ref{fig:observation-symbol-xml}, the latency varies for each application because each application has a different composition; as the number of symbols or XML tags in an application increases, it requires more time to launch the application.
For example, Cookie Run takes a long time for \emph{SO-Symbol Lookup}, which includes symbol search and PLT/GOT entry creation, because it should link a lot of SO files (\ie, symbols in SO pages) to launch the game.
But, Twitter shows the shortest time among the applications because it uses only two dynamic SO files.
Interestingly, the results on XML-UI conversion have a slightly different pattern from the above results; TikTok and WhatsApp reveal a long conversion time even though they spend a short time for \emph{SO-Symbol Lookup}.
\myrevcolorb{We also explored how SO file and XML-UI processing affect the application response/launch time.} These two costs \myrevcolorb{were the sole factors} affecting the application response time. On the other hand, in case of the application launch time, there are other factors (e.g., process initialization, user environment setting, user layout initialization, and update check) that can be explored, but we confirmed that they are negligible while analyzing our evaluation results.

These observations and experimental results motivate us to explore the design of SWAM.

%% file: section/40design.tex
To alleviate the overheads while preserving the benefits of existing schemes (ZRAM, Swap, LMKD, and OOMK), we propose a new memory monitor and management scheme, SWAM, for modern mobile devices.
The design goal of SWAM is to (1) reduce the launch time and response time of applications by orchestrating the behaviors of ZRAM and Swap mechanisms, (2) efficiently reclaim the memory space occupied by applications based on both the process attributes and access patterns, and (3) postpone killing applications as long as possible while optimizing it with the help of launch time estimation. 
\begin{figure}
  \centering
  \includegraphics[width=1.0\columnwidth]{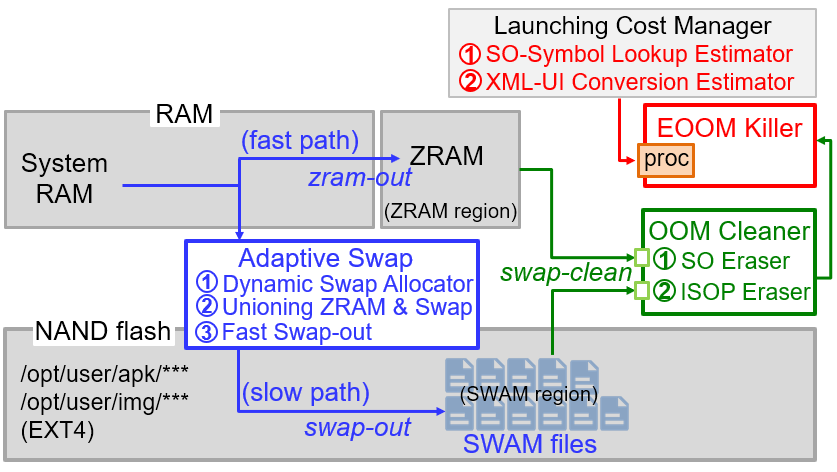}
 % \vspace{0mm} %Put here to reduce 
  \caption{System overview of SWAM}
  \label{fig:swam-overview}
%  \vspace{1mm} %Put here to reduce line intervals
\end{figure}
%~\\
In order to introduce the concept of SWAM before going further, Figure~\ref{fig:swam-overview} shows the architectural overview of SWAM, which is composed of \emph{Adaptive Swap}, \emph{OOM Cleaner}, and \emph{EOOM Killer}. 
As indicated in the figure, \emph{Adaptive Swap} simultaneously handles the \emph{zram-out/swap-out} procedure for achieving efficiency and better performance based on the characteristics of not only SO pages but also normal pages (see \textcolor{blue}{blue} lines). \emph{OOM Cleaner} secures enough free space for swap operations without killing applications (see \textcolor{black!30!green!70!}{green} lines). Finally, \emph{EOOM Killer} identifies victim applications that can be quickly re-launched based on the hints of the \emph{Launching Cost Manager} and kills then in the worst case (see \textcolor{red}{red} lines).

\subsection{Adaptive Swap}\label{ondemand-swap}
As mentioned in \S\ref{observ:so-lookup}, most mobile applications rarely share their SO pages with other applications and the existing swapping mechanisms \cite{swapping} do not consider such internal characteristics. However, the information on the SO pages which are shared across applications can open up an optimization opportunity for the swapping mechanism which determines the victim pages in memory. SWAM exploits such an opportunity and prioritizes non-shared SO pages for swap-out.
%~\\
~\\
\textbf{Dynamic Swap Allocator}: Unlike the traditional swap mechanisms, \emph{Adaptive Swap} can dynamically adjust the capacity of storage-class swap space in order to let the application's data be kept in the swap space as long as possible.
Besides, it can postpone the traditional OOMK operation, which terminates applications for memory reclamation, because it is triggered only when the amount of swap space is eventually insufficient; with \emph{Adaptive Swap}, we can adjust the amount of swap space suitably and dynamically.
\emph{Adaptive Swap} enables file systems to allocate swap space on-demand.
If an application requires more memory space than the available one, \emph{Adaptive Swap} scans the existing SO pages in memory and move the data in the SO pages into the files newly created in ordinary file system (\emph{\myrevcolorb{we} call them \textbf{swam files}}). 
Afterward, \emph{Adaptive Swap} handles the swap space accommodated in the files using normal file operations such as read(), write(), and unlink().

In addition, \emph{Adaptive Swap} uses a strict priority policy to carefully select the victim SO pages for swap-out. 
The priority level of each SO page can be classified into the following four categories according to the access or reference history of the SO page, which will be used to determine the victim SO page that is not likely to be reused in the future; (1) the SO pages that have not been accessed recently (during the predefined time interval), (2) the SO pages that have the reference count smaller than the predefined threshold, where the reference count means the number of processes sharing the SO page, (3) the SO pages that have been swapped out but now loaded in memory by swap-in operation, and (4) finally, among the remaining SO pages, the SO pages that are using larger amount of memory space.
Meanwhile, if a page fault occurs after swap-out, \emph{Adaptive Swap} issues a swap-in operation on the corresponding swam files, and it triggers unlink() operation when the file becomes empty by a series of swap-in operations.
Finally, if it is unable to obtain sufficient memory space after swapping out all the SO pages in memory to the \emph{swam files}, \emph{Adaptive Swap} begins scanning the normal pages (\eg, heap, stack, and shared memory) even though they have a high access frequency and swaps them out in the same manner as stated above.
%~\\
~\\
\textbf{Unioning ZRAM and Swap Mechanism:} In some cases, the operations in \emph{Adaptive Swap} can become the performance bottleneck because they employ the storage interface between the file system and the underlying storage devices.
To solve this issue, \emph{Adaptive Swap} provides two independent swap paths: (1) the (traditional) slow path that uses storage devices based on the Swap mechanism for a SWAM region and (2) the (accelerated) fast path that uses memory with ZRAM mechanism for a ZRAM region.
The slow path is designed on the dynamic swap allocator where the capacity of swap space in the underlying storage device can be dynamically adjusted.
Meanwhile, since even the fastest storage devices (\eg, eUFS and eMMC) are slower than the DRAM memory, the fast path is implemented for guaranteeing short latency by following the design principle of ZRAM.

When the available memory space is insufficient, \emph{Adaptive Swap} determines the swap path for memory reclamation by examining the information on the victim page.
First, if an SO page is selected as a victim to be swapped out, \emph{Adaptive Swap} counts how many applications share the SO page and then triggers the swap-out operation through the slow path in case of no sharing (\eg, application's SO pages). 
Otherwise, it opens the fast path to perform the swap-out (\ie, zram-out) of the victim SO page (\eg, platform's SO pages) into memory.
Second, if one of the normal pages is selected as a victim, it identifies whether the application owning the page is response-time-critical or not.
If the application is marked as time-critical in advance by manufacturers, the victim page is swapped out into memory via the fast path. Otherwise, the page is swapped out via the slow path.
Note that \emph{Adaptive Swap} first selects SO pages as victims to be swapped out until treating the whole SO pages in memory, and thus, the normal pages are rarely swapped out. 
The reason behind such priority assignment is that \emph{Adaptive Swap} was designed based on the observations mentioned in \S\ref{observ:so-lookup}.
%~\\
~\\
\textbf{Enabling Fast Swap-out}: Increasing or decreasing the unit size of unmap() operation has not been considered as a positive or negative factor in conventional swap-out mechanisms; the existing swap mechanisms have used the unit size of a fixed granularity.
Unfortunately, the smaller unit size may have a detrimental effect on the swap-out time because it increases the possibility of memory reclamation being delayed due to the interference with high-priority operations \cite{rtss17nasri}.
Especially, there may be a non-negligible delay in reclaiming a large amount of memory space.
To accommodate more adaptable swap-out, we propose a new unmap policy that enables rapid memory reclamation by dynamically modifying the unit size of the unmap() operation.
\vspace{-3mm} %Put here to reduce
\begin{algorithm}
\fontsize{9pt}{9pt}\selectfont
\algsetup{linenosize=\tiny}
\caption{Memory unmap to speed up swap-out}
\label{fig:fast-unmap-swapout}
\SetAlgoLined
\mbox{unmap( ):}
\Begin{
%// Start unmapping virtual memory address spaces \\
mem\_unmap\_unit = /proc/vm/dyn\_unmap\_sz; \mycircle{\scriptsize{1}} \\
Try mutex lock; \\
\While{memory region is not unmapped}{
  Task is uninterruptible (wait for completion); \\
  Unmap (mem\_unmap\_unit); \mycircle{\scriptsize{2}} \\
  Task is interruptible; \\
  \If{a higher priority task exists \mycircle{\scriptsize{3}}}{
       Get CPU resources; \\
       Run a scheduler (a higher priority task); \mycircle{\scriptsize{4}}\\
  }
}
Do mutex unlock; \\
%// End unmapping virtual memory address spaces \\
} 
\end{algorithm}
\vspace{-3mm} %Put here to reduce
To further understand our policy, we present its pseudocode in Algorithm \ref{fig:fast-unmap-swapout}.
We adopt the \emph{/proc} interface to quickly transfer and modify the unit size (\ie, mem\_unmap\_unit) of the unmap mechanism on-the-fly (\mycircle{\scriptsize{1}}).
The unit size is passed to unmap() operation to reset the mapping information of the virtual pages that were not swapped out (\mycircle{\scriptsize{2}}).
We next check whether there are a higher priority processes or not (\mycircle{\scriptsize{3}}) and if so, one of the existing higher priority processes will be instantly serviced (\mycircle{\scriptsize{4}}).
Due to the fact that the granularity of unmap() can be adjusted at runtime, it is possible to improve the response time of the swap-out operation by minimizing the number of check operations (\mycircle{\scriptsize{3}}).
For instance, let us assume that a swap-out process tries to free approximately 100 MB of memory space.
The traditional approach using unmap() with fixed granularity should conduct the operation that verifies the existence of higher priority processes (\mycircle{\scriptsize{3}}) up to 3,200 times in the case that the fixed granularity is 32 KB.
On the other hand, our policy triggers the operation (\mycircle{\scriptsize{3}}) only up to 25 times by setting the unmap granularity to 4 MB in advance.
Of course, as increasing the granularity, the unmap() operation might require more time to complete.
However, we believe that such delay has little impact on the scheduling of high-priority processes because they can be scheduled on other CPU cores.

%\subsection{Out-of-Memory (OOM) Cleaner}\label
\subsection{OOM Cleaner}\label{oom-cleaner}
SWAM should manage swap space in both memory and storage devices to assist \emph{Adaptive Swap}.
To elaborately manage the swap space, we design \emph{OOM Cleaner} that is composed \emph{Shared Object (SO) Eraser} and \emph{Idle Swapped-Out Page (ISOP) Eraser}.
SO Eraser and ISOP Eraser share the same goal of reclaiming swap space, but they have different target and policy.
First, SO Eraser targets SO pages on memory, and thus its reclamation operation is delayed as long as possible to allow applications to rapidly access the SO-symbols. 
Meanwhile, ISOP Eraser aims to reclaim \emph{swam files} for minimizing the delay involved in the process of making free rooms in swap space and to secure free space in storage devices in advance.
Therefore, ISOP Eraser periodically runs to remove \emph{swam files} accommodated in storage devices.
Note that \emph{OOM Cleaner} is orthogonal to the existing killing mechanisms in that it reclaims only the swap space that is related to SO pages (\ie, \swapclean) by exploiting the characteristics of the pages obtained from our observation (\eg, non-shared SO pages) instead of freeing all the memory space of an application at once.
\myrevno{\#1}\myrevcolor{\emph{OOM Cleaner} selects victim SO pages for swap-out in the order of access count and sharing count; the access count is the primary criterion and the sharing count is the tiebreaker.} If anonymous pages such as heap, stack, and shared memory are selected instead of the SO pages as victim pages, user response time may negatively be affected because these pages are accessed more frequently by applications.
% zram = compressed memory-backed swap = Compressed ZRAM Swap device
%~\\
~\\
\textbf{SO Eraser}: If the available memory space is insufficient even with the \emph{Adaptive Swap} enabled, \emph{OOM Cleaner} begins scanning the SO pages that reside in the ZRAM space, and reclaim the SO pages which are least shared by the applications in the system. Note that, the \swapclean, which reclaims the ZRAM space occupied by SO pages, is straightforward and less expensive than the aforementioned conventional swap-out methods because it does not require additional I/O operations.
Unfortunately, if an application calls one of the symbols belonging to the SO pages that were already reclaimed, it incurs the overhead of building up the entries for PLT and GOT, as explained in \S\ref{observ:so-lookup}, so that the page fault handler can reload the SO pages.
%~\\
~\\
\textbf{ISOP Eraser}: Since the SWAM region, unlike the ZRAM region, can be dynamically expanded, there is no need to remove the SO page using the ``least shared" strategy. Instead, \emph{ISOP Eraser} removes SO pages that have not been swapped-in for a long time to minimize the space waste in the SWAM region. To avoid keeping no-access swam files in the SWAM region, we designed \emph{ISOP Eraser} that removes the \emph{swam files} containing SO pages (\ie, \swapclean) in the storage devices (in LRU order) when the capacity of the storage device is insufficient or when the files are not used with a swap-in operation for a long time whose thresholds are set in advance. 
To do so, \emph{ISOP Eraser} runs periodically at the pre-defined interval, which can be adjusted to a shorter interval when the amount of available storage space falls below the pre-defined threshold.

%\subsection{Enhanced Out-of-Memory (EOOM) Killer}\label{eoom-killer}
\subsection{EOOM Killer}\label{eoom-killer}
Memory space pressure has a detrimental effect not only on the user experience (\eg, launch time and response time) but also on the usage of hardware resources (\eg, CPU cycles, storage devices, and power).
In this case, eliminating some of the running applications may be the best strategy to alleviate the stress by reclaiming a substantial amount of memory space at once, even if it involves a lengthy rebuilding time to allocate memory space and load the application's associated data again  \cite{android-lmkd, android-app-startup-time, access20-selective-swap-kill, mobisys12-fast-app-launching}.
To pursue this principle, \emph{EOOM Killer} inherits the functions of the traditional OOMK on the mobile device.
Like OOMK, \emph{EOOM Killer} eventually comes into play when the available memory space is insufficient despite of running \emph{Adaptive Swap} and \emph{OOM Cleaner}; of course, the operations of \emph{OOM Cleaner} and \emph{EOOM Killer} can be overlapped.
To extend the process of selecting a victim in OOMK, \emph{EOOM Killer} leverages the relaunch cost of SO-symbol lookup and XML-UI conversion, which is passed through a communication channel between the user and kernel-space (\eg, the \emph{/proc} interface).
Since the relaunch cost differs widely depending on the composition of applications, it is determined by combining the results of \emph{SO-Symbol Lookup Estimator} and \emph{XML-UI Conversion Estimator}, obtained when an application is launched for the first time.
%~\\
~\\
\textbf{SO-Symbol Lookup Estimator}: According to our observation in \S\ref{observ:so-lookup}, popular applications mostly have SO files included \cite{android-ndk},
which incur an extra overhead of SO-symbol searching to construct PLT and GOT. 
We can improve the process of selecting a victim application based on the overhead; \ie, we can identify which application requires more time for relaunch by calculating the cost of such run-time overhead.
For example, if an application needs a relatively longer relaunch time, \emph{EOOM Killer} gives positive chances for it to reside in memory as long as possible.
The SO-symbol lookup cost can be estimated by the following equation:
\setlength{\abovedisplayskip}{0pt}
\setlength{\belowdisplayskip}{1pt}
\begin{align*}
&Ts = Lookup\;time\;for\;a\;symbol &&\\
&Tr = Time\;to\;relocate\;a\;memory\;address\;of\;a\;symbol &&\\
&Tl = Loading\;time\;of\;an\;SO\;file &&\\
&T = (\sum_{for\;each\;symbol} (Ts + Tr)) + Tl
\end{align*}
Here, \emph{T} is the total symbol lookup latency; the total symbol lookup spends more time as the number of symbols for lookup increases.
%~\\
~\\
\textbf{XML-UI Conversion Estimator}: \emph{EOOM Killer} tries its best effort to select a victim while minimizing the impacts on user experience. 
As applications' user interfaces get more complex, the XML processing cost increases, and \emph{EOOM Killer} is one of the complementary methods for reducing XML-UI conversion overhead; our observations in \S\ref{observ:xml-cost} led us to this conclusion.
The estimator monitors the XML processing cost in loading phases: (1) the time required to alter the UI layouts and (2) the time required to render the contents on the screen. 
The cost provides a hint as to which application leads to more overhead, together with the lookup cost.

%% file: section/70evaluation.tex
In this section, we evaluate SWAM with the popular mobile applications so as to answer the following research questions: (\textbf{RQ1}) \emph{how does each individual effort in SWAM help with taking care of the available memory space} and (\textbf{RQ2}) \emph{how well does SWAM perform on the whole with modern mobile devices?}

\subsection{Experimental Setup}\label{eval_setup}
In our experiments, we target Android platform because it is one of the widely used mobile platforms, where its memory shortage issue has detrimental effects on the user experience in that the device is dedicated to a user and foreground applications in the device directly communicate with the user. However, SWAM's three major components are built on top of the Linux kernel and the native C/C++ SO library, and so they are completely compatible with and easily portable to other Linux distributions.
\vspace{0mm}
\begin{table}[h!]
\caption{Applications and automated user interaction}
% The standard font size switches are: \tiny, \scriptsize, \footnotesize, \small, \normalsize, \large, \Large, \LARGE, \huge, and \Huge.
\footnotesize
\begin{minipage}{\columnwidth}
\begin{center}
\setlength\tabcolsep{4.5pt} % default value: 6pt
\begin{tabular}{l|l|l}
\hline
% \mycirclethree{\scriptsize{1}}, \mycirclethree{\scriptsize{2}}
\textbf{Category} & \textbf{Foreground applications} & \textbf{Auto user inputs}  \\
\hline
Media     & Youtube, Netflix, Podcast  & Watch videos            \\
Messaging & Facebook, Twitter, TikTok, & Browse and read posts   \\
          & Skype, WhatsApp, Viber     &                         \\ 
News      & BBC News, NewYork Times    & Browse and read articles \\
Game      & CookieRun, AngryBird       & Play a stage             \\
Internet  & Chrome, Firefox            & Browse and read posts    \\
\hline
\end{tabular}
\end{center}
 \myrevcolor{* Background applications: Media (Pinterest, Jellyfin, Kodi, Spotify), Messaging (Instagram, Telegram, Discord, Snapchat), Note Organizer (Evernote, Notion, OneNote, Colornote), Trip (Airbnb, Rentalcars, Skyscanner), Office (Dropbox, TeraBox, OneDrive, GoogleDrive), Game (Candy Crush Saga, Clash of Clans, Subway Surers), and News (Google News, Reddit, Flipboard).}
 \\
\label{table:eval-scenario}
\end{minipage}
\end{table}
\vspace{0mm} %Put here to reduce
~\\
\textbf{Hardware and Software}: We implemented the prototype of SWAM in Linux kernel version 5.10 and Android Open Source Project (AOSP) 12 \cite{android-aosp, android-factory-image}; the total of 7,250 lines of code were added or modified \cite{swam}.
We also used real world applications (top 15 applications from Google Play \cite{google-play}, as shown in Table \ref{table:eval-scenario}) that are frequently employed. The behavior of swapping and killing approaches significantly depends on memory capacity, and therefore, to investigate the possible impact of SWAM on limited hardware resources, \myrevcolorb{we performed the evaluation on a state-of-the-art high-end mobile device (\ie, Google Pixel 6) and an emulated low-end mobile device.}
\myrevcolorb{The high-end device is equipped with Octa-core ARM CPU, 8 GB memory, and 128 GB eUFS 3.1 storage device. We compared SWAM with two baselines where the original LMKD/OOMK is already enabled}; (1) \textbf{ZRAM} with 1 GB memory for swap space and \texttt{lz4} algorithm \cite{lz4} for (de)compression, and (2) \textbf{NAND-swap} with 3 GB swap partition reserved on the underlying storage device.
For the experiments on low-end device \cite{galaxy-a12}, we used the same high-end device while limiting its CPU, memory, and eUFS storage resources to Quad-core, 4 GB, and 64 GB, respectively; in this case, the configuration of the baseline was adjusted to 512 MB for \textbf{ZRAM} and 2 GB swap partition for \textbf{NAND-swap}.
We also modified about 20 lines of code in the multi-core scheduler and memory layout of the Linux kernel to enable the Google Pixel 6 device operate as a low-end device with limited system resources.
%~\\
~\\
\textbf{Methodology}: For a fair comparison, we selected widely used mobile applications as mentioned in Table \ref{table:eval-scenario}, and conducted experiments by following the four steps on both low-end and high-end mobile devices: (1) We installed the pre-selected 40 applications (15 real-world applications and 25 background applications) on the device to begin each experiment. (2) We set up the initial test environment by performing memory operations, which consume 256MB, to bring the memory pressure situation. (3) We used \emph{adb} and \emph{logcat} commands to collect evaluation results while performing automated tests with \emph{UI Automator} that emulates UI touches of users based on scripts. We performed the same automated tests (i.e., step 3) every day for 1.5 hours without a reboot process for 4 weeks. \myrevno{\#2}(4) Finally, we rebooted the device to remove any effects from the previous experiments and \myrevcolor{to reset the user configuration settings, which include notifications, background data limits, and permission restrictions}. Note that we repeated the evaluation steps \myrevcolorb{2--4} whenever changing evaluation configurations (i.e., NAND-swap, ZRAM, ZRAM/NAND-swap, and SWAM).

\myrevcolor{To imitate mobile users for realistic evaluation, we referred SIMFORM's premium user document \cite{simform21-private}}\footnote{\myrevcolor{Unfortunately, normal users cannot view this document. So, we cite the publicly accessible document \cite{simform21-public}.}} \myrevno{\#3}; this document covers users’ application preferences and usage patterns (e.g., the average number of installed applications on mobile phones and their average daily usage pattern). To the best of our knowledge, it is well known that SIMFORM provides good guidance on the pattern in Mobile \& CE area.
We set the test time to \myrevno{\#4}\myrevcolor{1.5 hours based on the description of "\textit{about 1 hour, 43 minutes a day}" in Section 2 of the reference \cite{simform21-public}; the test time is set conservatively to clearly confirm the benefits of SWAM.} Based on the description "\textit{an average person has 40 applications installed on his phone}" in Section 1 of the reference \cite{simform21-public}, we installed 15 foreground applications and \myrevno{\#5} \myrevcolor{25 background applications, which are ranked higher in popular application categories from Google Play Store \cite{google-play}.} At this time, we performed an automated test for 1.5 hours on mobile devices and afterwards put the devices in an idle state everyday. Even in the idle state, applications consume memory space because they periodically run in the background as explained in \S\ref{observ:bg-service} (Memory Space
Consumption). As a result, some applications may have to be killed over time for memory reclamation.
\vspace{0mm} %Put here to reduce
\begin{figure}
  \centering
  \includegraphics[width=0.99\columnwidth]{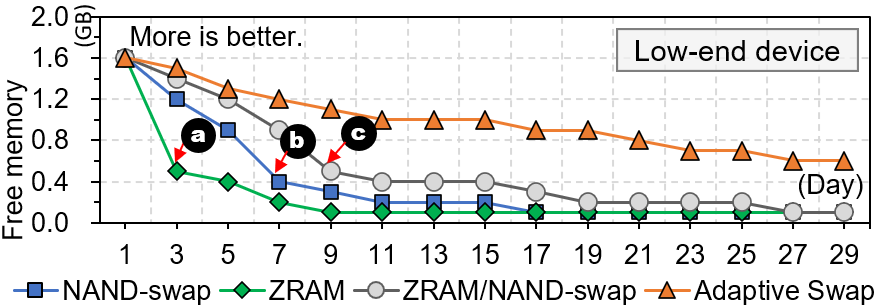}
  \caption{The average amount of free memory space}
  \label{fig:feature-adaptive-swap}
  \vspace{0mm} %Put here to reduce too much white space after your table
\end{figure} 
\vspace{0mm} %Put here to reduce
%~\\
%~\\
\subsection{Contributions of Each Component}\label{eval:feature-components}
First, to understand the impact of each component of SWAM, we conducted simple experiments that can figure out the effectiveness of each component on low-end mobile devices.
~\\
\textbf{Adaptive Swap}:
First, we confirm how much free memory space can be secured by enabling only the \emph{Adaptive Swap}.
Figure~\ref{fig:feature-adaptive-swap} traces the amount of free memory space of the four swap mechanisms: NAND-swap, ZRAM, ZRAM/NAMD-swap, and \emph{Adaptive Swap}.
%As mentioned in \S\ref{eval_setup} (Methodology), we repeated an experiment for 1.5 hours per day for 4 weeks.
Interestingly, we can see that the average amount of free memory space decreases significantly on the 3\textsuperscript{rd} day for ZRAM  (\mycircle{\scriptsize{a}}), on the 7\textsuperscript{th} day for NAND-swap (\mycircle{\scriptsize{b}}), and on the 9\textsuperscript{th} day for ZRAM/NAND-swap (\mycircle{\scriptsize{c}}).
These trends mean that applications quickly consume free memory space and they may negatively impact the performance of mobile devices.
On the other hand, \emph{Adaptive Swap} guarantees stable free memory space, at least 600 MB on average.
Such different results comes from the fact that \emph{Adaptive Swap} gracefully takes the benefits of the slow path by dynamically adjusting the swap space in the underlying storage device.
Thus, we additionally measured the storage utilization so as to verify how many \emph{swam files} are dynamically created and deleted to support the slow path in \emph{Adaptive Swap}. 
As a result, we found that the peak usage of the swap space reaches 6.2 GB at most on the 64 GB storage device.
\begin{figure}
  \centering
  \includegraphics[width=0.99\columnwidth]{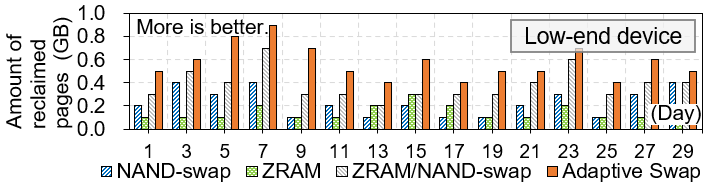}
  \caption{The amount of SO pages that are reclaimed by \emph{OOM Cleaner}}
  \label{fig:feature-oom-cleaner}
\vspace{1mm} %Put here to reduce
\end{figure}
~\\
\textbf{OOM Cleaner}:
Since \emph{OOM Cleaner} is an auxiliary component to secure swap space of SWAM, it can be enabled with the conventional swap mechanisms.
To clearly understand the effectiveness of \emph{OOM Cleaner}, we applied \emph{OOM Cleaner} with each of the four swapping mechanisms.
Figure~\ref{fig:feature-oom-cleaner} shows the total amount of SO pages that are reclaimed by \emph{OOM Cleaner}.
For the evaluation of NAND-swap, we modified \emph{ISOP Eraser} so as to periodically reclaim swap space at the granularity of SO page instead of swam file; NAND-swap handles swap-in/out in the page unit.
As expected, ZRAM shows less efficiency in reclaiming SO pages compared to other mechanisms because it has no swap space in storage devices.
In other words, ZRAM can only reclaim SO pages allocated in memory on demand by \emph{SO Eraser}, and thus it has less chance of reclamation compared with the other schemes.
Meanwhile, ZRAM/NAND-swap shows better adaptability than ZRAM or NAND-swap because \emph{SO Eraser} and \emph{ISOP Eraser} independently reclaim swap space on memory and storage devices, respectively.
In this case, when the amount of fixed swap space of NAND-swap is insufficient, SO pages may be reclaimed by \emph{ISOP Eraser}.
\emph{OOM Cleaner} with \emph{Adaptive Swap} achieves the highest efficiency compared to conventional swap schemes; it comes from the fact that \emph{ISOP Eraser} has more opportunities to secure the swap space than NAND-swap because the swam files can dynamically be created by \emph{Adaptive Swap}. 
Note that, swam files are created by \emph{Adaptive Swap} and they are deleted by the background operations of \emph{ISOP Eraser}.
%~\\
\begin{figure}
  \centering
  \includegraphics[width=0.99\columnwidth]{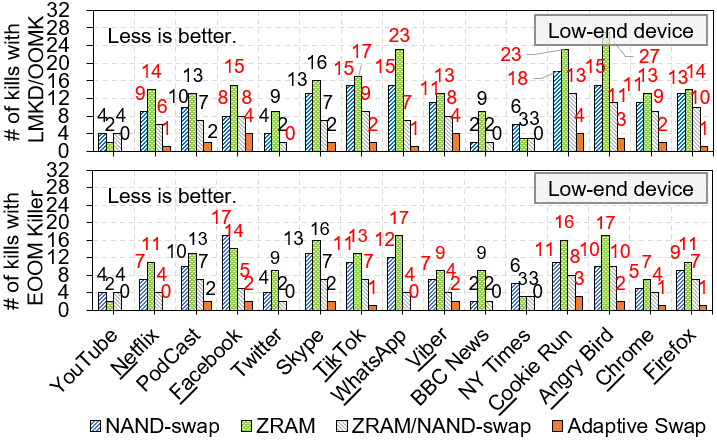}
  \caption{The number of kills with LMKD/OOMK (above) and \emph{EOOM Killer} (below)}
  \label{fig:feature-eoom-killer}
\vspace{1mm} %Put here to reduce 
\end{figure}
%~\\
~\\
\textbf{EOOM Killer}:
Now, let us add \emph{EOOM Killer} to the four swap mechanisms to see how long the applications can continue to run without being terminated. Figure~\ref{fig:feature-eoom-killer} plots the number of times each application was forcibly terminated by LMKD/OOMK and \emph{EOOM Killer}. In this figure, the number on each bar means how many times each application was killed and re-launched during running the experiments. As expected, \emph{EOOM Killer} contributes on the number of terminations of all the applications for each swap mechanism. For example, The number of terminations for Cookie Run is decreased from 18 to 11 in NAND-swap, from 23 to 16 in ZRAM, from 13 to 8 in ZRAM/NAND-swap, and from 4 to 3 in \emph{Adaptive Swap}.

As illustrated in Figure~\ref{fig:feature-eoom-killer}, \emph{EOOM Killer} kills fewer foreground applications compared to OOMK. However, it doesn't mean another process is forcibly killed instead. If \emph{EOOM Killer} has to kill an application for memory reclamation in an extreme memory shortage situation, it just prefers to kill applications having a short launch time; \myrevcolorb{in} our experiments, the killed applications were mostly background applications. This is because, unlike foreground applications, background applications do not require GUI operations and user configuration settings for user interaction.

Meanwhile, \emph{Adaptive Swap} rarely triggers LMKD/OOMK or \emph{EOOM Killer} compared with the other swap mechanisms because it stably guarantees the available memory space for applications, as shown in Figure~\ref{fig:feature-adaptive-swap}. In addition, \emph{EOOM Killer}, when combined with \emph{Adaptive Swap}, records the lowest number of terminations for all applications. For example, it kills Cookie Run only 3 times by considering the expensive re-launch time on victim selection. We believe that these positive effects can lead it to the higher level of user experience.

\subsection{Integrated Evaluations}\label{eval:ux-all}
This section describes the experimental results of the integration test. SWAM is compared to the three representative solutions: NAND-swap, ZRAM, and ZRAM/NAND-swap.

To clearly measure the effectiveness of each component, we injected the \emph{logcat} command, which prints out log messages from the system, including memory traces, in the code at the triggering point of each component and we connected mobile devices with a terminal machine using Android Debug Bridge (ADB). Then, we performed evaluation tests while collecting the raw data individually generated by each component through
ADB on-the-fly. In other words, the evaluation results mentioned in \S\ref{eval:ux-oomk-freq} (\# of killed applications) and \S\ref{eval:app-performance} (the application launch time and response time) are based on the raw data collected from the final integration test. In addition, we performed extra experiments where we
separately turned on each component to independently collect the message without any interference from other components.
\begin{figure}
  \centering
  \includegraphics[width=0.99\columnwidth]{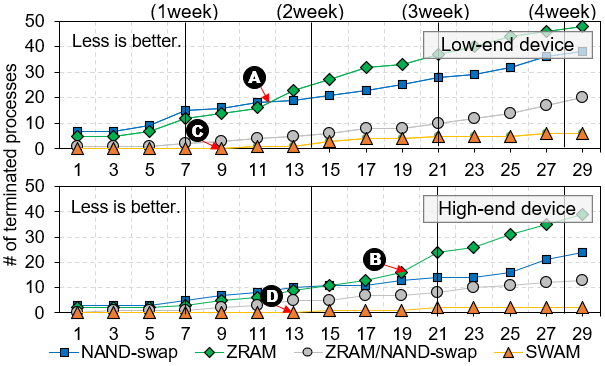}
  \caption{The number of accumulated OOMKs on low-end (above) and high-end (below) mobile devices}
  \label{fig:ux-oomk-frequency}
\vspace{1mm} %Put here to reduce too much white space after your table
\end{figure} 
%~\\
\subsubsection{Number of Forcibly Killed Applications} \label{eval:ux-oomk-freq}~\\
We performed experiments on the number of times the LMKD/OOMK operations are executed, to see how much the SWAM improves the frequency of the operations that forcibly kill running applications regardless of the user's intention. 
Figure~\ref{fig:ux-oomk-frequency} shows our experimental results on both the low-end and high-end devices using an interaction scenario. 
In this scenario, an automated user input test was conducted by launching the applications listed in Table \ref{table:eval-scenario} and simulating intensive user inputs for a duration of 1.5 hours; some processes may be terminated due to memory pressure regardless of their behavior during testing.
We relaunched the killed applications the next day to clearly understand how many processes are frequently killed every day.
Of course, since background services and surviving applications consume memory space for the remaining 22.5 hours, they can give a negative effect on the available memory space.

In this figure, ZRAM collapses, losing its benefits after 12\textsuperscript{th} day on the low-end device (\mycircle{\scriptsize{A}}) and 19\textsuperscript{th} day on the high-end device (\mycircle{\scriptsize{B}}).
The reason behind this is that the swap space on memory becomes full, and thus LMKD/OOMK runs to aggressively secure sufficient memory space.
ZRAM/NAND-swap that adopts both ZRAM and NAND-swap mechanisms shows stable patterns like SWAM, but it progressively increases the number of applications killed by LMKD/OOMK over time.
Meanwhile, SWAM shows an ideal pattern where the killing operations to reclaim the memory space do not appear until 9\textsuperscript{th} day~(\mycircle{\scriptsize{C}}) and 13\textsuperscript{th} day~(\mycircle{\scriptsize{D}}) in low-end and high-end devices, respectively.
In SWAM, only 2 applications (on the high-end device) and 6 applications (on the low-end device) were killed by \emph{EOOM Killer} during 4 weeks.
To gain better understanding on the contributions, we performed the same analysis on the effect of each component of SWAM as we did in the unit test.
We found the improvement of SWAM comes from Adaptive Swap (61\%), OOM Cleaner (36\%, \emph{SO Eraser 27\% + ISOP Eraser 9\%}), and EOOM Killer (3\%), which is obtained by analyzing the contribution of each SWAM component on memory space securement.
\begin{figure} %[!ht]
  \centering
  \includegraphics[width=1.00\columnwidth]{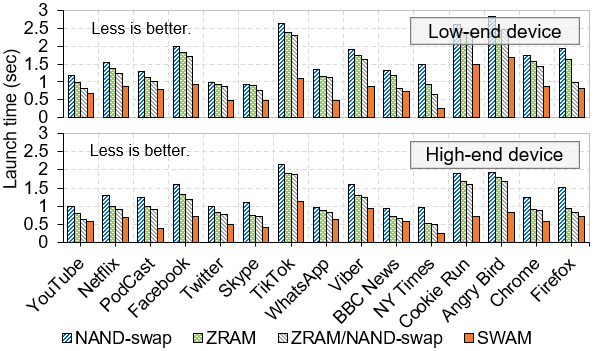}
  \caption{Launch time of applications on low-end (above) and high-end (below) mobile device}
  \label{fig:ux-app-launch-time}
\vspace{1mm} %Put here to reduce too much white space after your table
\end{figure}

\subsubsection{Application Performance} \label{eval:app-performance}~\\
The ultimate goal of SWAM is to ensure rapid interaction between a user and a mobile device.
To show the contribution of SWAM on the latencies, we assessed both application launch time and response time which are measured for user interactions with each application.
%~\\
~\\
\textbf{Application Launch Time}: 
We observed how much the SWAM reduces the launch time of applications on both high-end and low-end devices. 
In general, the launch process consists of a series of steps such as data loading, caching, and initialization; we already investigated some of the initialization costs (\ie, \emph{SO-Symbol Lookup} and \emph{XML-UI Conversion}) in \S\ref{observ:so-xml-cost}.
Figure~\ref{fig:ux-app-launch-time} shows the total time elapsed to launch each application. The conventional approaches reveal different results according to memory capacity (i.e., high-end and low-end). \myrevno{\#6}\myrevcolor{However, SWAM generates similar results on both devices because it efficiently secures sufficient free memory space to launch applications by releasing superfluous space in advance.} In other words, SWAM can reduce the number of swap-in/out operations while launching applications.
As shown in Figure \ref{fig:ux-app-launch-time}, SWAM improves the launch time by 38\% (on the low-end device) and 36\% (on the high-end device) compared to the two baselines and ZRAM/NAND-swap.
This enhancement is the result of effective orchestration of the three components of SWAM, especially accomplished by two significant factors.
One is that \emph{OOM Cleaner} secures enough free memory space to launch applications by releasing the space that is unlikely to be used or shared in advance.
This advantage can help accelerate the launch time of applications that require a significant amount of memory at startup and lower the number of swapping operations required during a series of startup procedures. For instance, on Netflix, which requires approximately 604 MB of memory to launch, SWAM improves its launch time by 29\% when compared to ZRAM/NAND-swap on a low-end device.
\begin{figure} %[!ht]
  \centering
  \includegraphics[width=1.00\columnwidth]{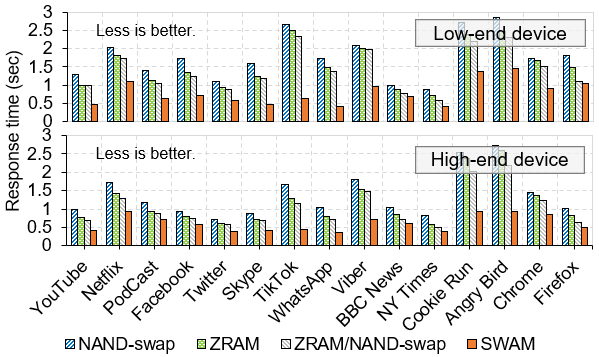}
  \caption{The average response time of applications on low-end (above) and high-end (below) mobile device}
  \label{fig:ux-app-response-time}
  \vspace{1mm} %Put here to reduce too much white space after your table
\end{figure}
The other is that, because SWAM permits \emph{Fast Swap-out}, the launch process is rarely interrupted in the middle of a swapping operation that reclaims a large amount of memory space. 
Finally, we also obtained the contribution of each SWAM component on the launch time of applications: Adaptive Swap (53\%), OOM Cleaner (39\%, \emph{SO Eraser 34\% + ISOP Eraser 5\%}), and EOOM Killer (8\%).
~\\
\textbf{Application Response Time}: 
The response time may directly be dependent on whether the application is killed or not, as it may involve the above-mentioned launch time. 
Figure~\ref{fig:ux-app-response-time} shows the average response time of the 15 popular applications while using the aforementioned interaction scenarios.
In the figure, SWAM has a noteworthy response time in all applications; it is the most effective method in that it reduces application response time by 44\% (on the low-end device) and 41\% (on the high-end device) compared to the two baselines and ZRAM/NAND-swap.
This startling result is achievable because popular applications are seldom chosen for space reclamation with the practical interaction scenario.
Therefore, the applications can instantly react to users almost at all times.
As a result, SWAM surpasses NAND-swap by two times in some applications, including YouTube, Skype, TikTok, WhatsApp, Viber, Cookie Run, and Angry Bird.
To ascertain the improvement in detail, we also tracked contribution of each component of SWAM on the response time and discovered that it comes from Adaptive Swap (43\%), OOM Cleaner (51\%, \emph{SO Eraser 38\% + ISOP Eraser 13\%}), and EOOM Killer (6\%).
~\\
~\\
\myrevno{\#7}\myrevcolor{Finally, to confirm the overhead, we analyzed the extra behaviors of SWAM by comparing it with the existing system in detail. Then, we found that SWAM shows very small overhead, which comes from the additional I/O operations for storage-class swap in \emph{Adaptive Swap} (1\% slower), the extra operations for removing SO pages in \emph{OOM Cleaner} (3\% slower), and the auxiliary costs for estimating execution time of SO-Symbol lookup and XML-UI processing (2\% slower). But, we believe this is negligible and the overhead is overshadowed by the performance advantage of SWAM (41\% faster application response time).}

%% file: section/75discussion.tex
SWAM dynamically adjusts the swap space in the file system, by increasing or decreasing the amount of available swap space.
Unfortunately, SWAM has an intrinsic constraint that precludes users from using the storage space consumed by \emph{\textbf{swam files}}.
When the amount of space occupied by \emph{\textbf{swam files}} increases, this issue can lead to new space competitions between users and SWAM.
For example, the file system possibly may not allow for \emph{Dynamic Swap Allocator} to add a new \emph{\textbf{swam file}} because of the peak utilization of the storage devices.
But, SWAM attempts to provide appropriate amount of free storage space in practical scenarios by orchestrating the functions of its three components.
%In the extreme case, since the \emph{\textbf{swam file}} space grows significantly, it may be challenging to create a user's file.
\begin{figure}
  \centering
  \includegraphics[width=0.99\columnwidth]{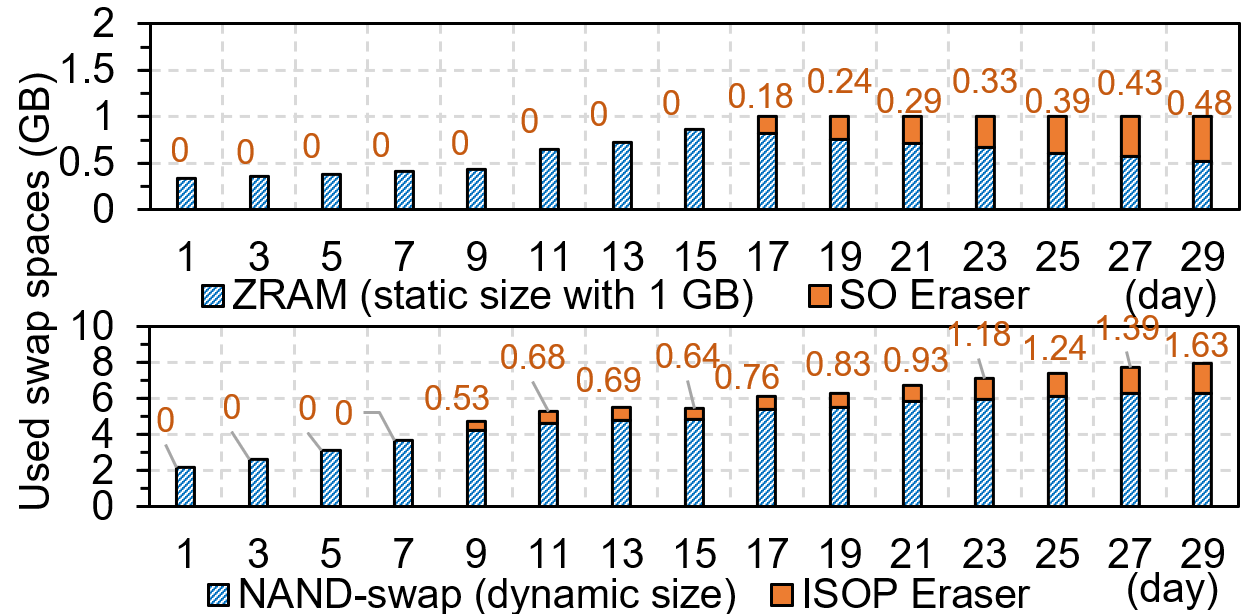}
  \caption{The swap space used on ZRAM (above) and NAND-swap (below). Numbers in orange color show how much the swap space shrinks by \emph{SO Eraser} and \emph{ISOP Eraser}.}
  \label{fig:discussion-swap-space}
  \vspace{1mm} %Put here to reduce
\end{figure} 
%~\\
Figure~\ref{fig:discussion-swap-space} depicts the amount of swap space that is reclaimed by \emph{SO Eraser} and \emph{ISOP Eraser}; they reclaim the swap space when the amount of swap space in use exceeds the ZRAM capacity or when \emph{\textbf{swam files}} are not utilized during a certain amount of time. Two configurations (\ie, the ZRAM capacity and the time period) can be adjusted considering the hardware characteristics of the mobile device and they can be defined by the manufacturer.
Our simple experimental results show that SWAM can safely conserve memory and storage consumption; \emph{SO Eraser} and \emph{ISOP Eraser} reduced the size of swap space by 0.48 GB and 1.63 GB, respectively. 
Theoretically, SWAM might consume a lot of storage space to secure free space on memory. However, as shown in Figure~\ref{fig:discussion-swap-space}, we believe that it is unlikely to occur in real-world situations because SWAM systematically reduces the possibility of the aforementioned issue by freeing storage space with \emph{ISOP Eraser} in advance.

Most mobile applications save memory space by using SO pages that share the memory space. Therefore, if applications are designed without SO pages, they may waste memory resources significantly. However, even in the case that the applications do not use SO pages, we believe some
components of SWAM (i.e., \emph{Dynamic Swap Allocator}, \emph{Fast Swap-out} module for \emph{Adaptive Swap}, and \emph{EOOM Kille}r) are still valuable. Also, when more and more SO pages are shared among the applications, SWAM becomes more important and efficient in that the number of swap in/outs is reduced by keeping most shared SO pages in memory.

% original: ... they reclaim the swap space which exceeds a time period or a storage limitation.

%% ###################### ENGLISH ######################

%% file: section/80relatedwork.tex
% https://struggler.tistory.com/m/426
% https://tex.stackexchange.com/questions/93797/how-to-rotate-text-inline
% https://workspace.google.com/marketplace/app/spreadlatex/218144906748
% 
% \begin{table}
%   \centering
%   \caption{Comparison of mobile optimization strategies}
%   \includegraphics[width=0.99\columnwidth]{img/relatedwork}
%   \label{fig:relatedwork}
% \vspace{0mm} %Put here to reduce too much white space after your table 
% \end{table}
% In this section, we discuss the literature most relevant to our work.
% In this section, we categorize recent research and engineering approaches aimed at securing the available memory space into swapping (e.g., RAM, NVRAM, and NAND flash) and killing strategies. 
%This section discusses swapping (\eg, NVRAM and NAND flash) and killing strategies, as well as responsiveness optimization techniques (\eg, RAM), most relevant to our work.
This section discusses swapping (\eg, RAM, NVRAM, and NAND flash) and killing strategies, relevant to our work.
%~\\
~\\
\textbf{RAM.} 
As mentioned before, the RAM-based swapping mechanism is one of the common approaches to speed up swapping operations. 
\emph{ZRAM} \cite{linux-zram} and \emph{Zswap} \cite{linux-zswap} are designed to use DRAM as their swap space instead of the storage devices to speed up swapping operations.
In addition, they efficiently save the available swap space by compressing pages to be swapped out and storing them in compressed format.
To further secure the available memory space, \emph{ezswap} \cite{access19-ezswap} allows to swap out not only anonymous pages but also pages mapped to the files.
%~\\
~\\
\textbf{NVRAM.} 
NVRAM (\eg, PRAM, RRAM, and STT-MRAM) is considered to be suitable for swap devices because of its positive features such as short latency, low power consumption, and high density.
Unlike SWAM, \emph{K. Zhong} \cite{emsoft14-nvram} proposed an NVRAM-based swap technique for fast swapping.
However, because NVRAM has a capacity problem in comparison to NAND flash, it is still much more expensive to replace NAND flash with NVRAM due to its higher manufacturing cost.
Therefore, the industry is still adopting NAND flash-based swap devices, while using compressed memory as a swap cache when necessary \cite{linux-zswap, ram-plus, ram-plus-ac, ram-expansion-oppo}.
%~\\
~\\
\textbf{NAND flash.} Many efforts have been made for efficient swapping that can keep the state of running applications on NAND flash storage devices.
\emph{Flash-aware Linux Swap} \cite{tce12-flash-swap} controls I/O patterns to mitigate the endurance issue of the flash storage devices. 
\emph{Marvin} \cite{atc20-marvin} modifies Android Run-Time (ART) to make a set of efforts for selecting the pages unused for a long time as victims. 
Meanwhile, \emph{SmartSwap} \cite{dac17-smartswap} predicts which applications will not be used in the future and reclaims the pages belonging to the applications in the swap-out phases.
%~\\
~\\
\textbf{Killing Techniques.} 
The traditional low memory killer identifies victim processes based on their priority and the amount of pages they have. 
\emph{Android LMKD} \cite{android-lmkd, lwn12-us-lmkd, mobisys12-fast-app-launching, iwcmc17-opt-lmk-ml} uses the \textit{OOM score} of applications to forcefully kill some processes in the user space. 
\emph{SmartLMK} \cite{tecs16-smart-lmk} and \emph{POA} \cite{tecs14-person-opt} supply some statistics and application usage patterns indicating the launch time and the frequency of application activations to the operating system as a hint, so that the operating system can efficiently identify and eliminate victim processes.
If the available memory space is not sufficient despite of running LMKD operations, \emph{OOMK} \cite{lwn10-another-oomk} is triggered with its heuristic policy to avoid system shutdown.
However, the LMKD/OOMK approach irritates mobile device users by forcibly closing user applications regardless of the user's intention.

To prevent the coercive termination of processes by the kernel-mode killer, \emph{Hybrid swapping} \cite{access18-hybrid-swap} and \emph{SEAL} \cite{tcaics20-seal} offer a two-level swapping technique based on ZRAM and NAND-swap. However, these solutions concentrate on application launch time and do not address application response time or dynamic swap space.

Meanwhile, \textit{App hibernation} \cite{app-hiber} which is suggested in Android 12 has a similar mechanism to SWAM in that it can take out memory space from certain applications. But, unlike SWAM, it focuses on suspending dormant applications that has been idle for a long time. On the other hand, SWAM tries to secure free memory space by reclaiming even the SO pages of recently executed applications without suspending the applications (see \S\ref{sec:design}). Therefore, SWAM and \textit{App hibernation} may target different applications, and we believe that both can collaboratively operate.
~\\
~\\
Conventional studies have focused on pre-processing techniques that should be performed in advance to optimize the application performance on the mobile platform. On the other hand, our work accelerates application launch time and response time by combining the swapping and killing operations concinnously to enable fine-grained and efficient memory management on mobile devices.

%% file: section/90conclusion.tex
We carefully revisit the conventional memory management techniques in mobile platforms.
We highlight the structural limitations of the traditional swapping and killing mechanisms which operate independently, and propose a state-of-the-art memory management technique that complements the traditional schemes.
To improve application responsiveness even in low memory conditions on mobile devices, we not only preserve the process state by trying to eliminate only the SO pages that have been inactive for a long time, but also reduce the frequency of swap-in/outs considerably.
As a result, even when a mobile device is under memory pressure, it may provide faster application launch time (36\% faster) and faster response time (41\% faster) than the conventional schemes, which shows significant positive impact on user experiences.

%% ###################### ENGLISH ######################

%% file: section/95ack.tex
% We appreciate the insightful comments and feedback from the anonymous reviewers and shepherd. This work was partly supported by the Ministry of Science and ICT (NRF-2017R1A2B3004660) and Samsung Research (RAJ0121ZZ-32RF).

% 일단 과제 카테고리는 "기초연구사업 - 기본연구" 입니다
% 
% DCLAB 과제가 속한 국가과학기술표준분류는
% 정보/통신 - 정보이론 - 오퍼레이팅 시스템 (가중치: 60%)
% 정보/통신 - 소프트웨어 - S/W 솔루션 (가중치: 30%)
% 정보/통신 - 소프트웨어 - System Integration (가중치: 10%)
% 

% Geunsik Lim, ## 2023년도 과제번호: 
% * 연구과제 정보: S/W Dev. Assistant 및 품질 분석 도구 개발 (RAJ0123ZZ-25RF)
% * 착수 DR: 2023.02.28
% * 완료 DR: 2023.12.31
% 
% * PLM 과제 관리: http://splm.sec.samsung.net/

% full statements
%\myrevcolor{We would like to thank the anonymous reviewers and the shepherd for their valuable comments and suggestions. We would like to further thank Seungkeun Lee, Jaewon Kim, Chanwoo Choi, Inki Dae, HyoYoung Kim, Dongseob Park, Jonggyu Park, and Heekuk Lee for their insightful discussions and feedback. This work was partly supported by the National Research Foundation of Korea (NRF) grant funded by the Korea government (MSIT) (No. NRF-2022R1F1A1070065) and Intelligent Dev. Assistant \& Quality Tool Development (RAJ0121ZZ-32RF) of Samsung Research.}

% abbreviation and short statements
\myrevcolor{We appreciate the insightful comments and feedback from the anonymous reviewers, the shepherd, Jonggyu Park, Seungkeun Lee, Jaewon Kim, Chanwoo Choi, Inki Dae, HyoYoung Kim, Dongseob Park, and Heekuk Lee. This work was partly supported by the NRF grant funded by the Korea government (MSIT) (No. NRF-2022R1F1A1070065) and Quality Tool Lab of Samsung Research (No. RAJ0123ZZ-25RF).}

% abbreviation and short statements
%\myrevcolor{We would like to thank the anonymous reviewers and the shepherd for their valuable comments and suggestions.} 
%\myrevcolor{We would like to further thank Blind01, Blind02, Blind03, Blind04, Blind05, Blind06, and Blind07.}
%\myrevcolor{This work was partly supported by Blind-000000A 00000A 000A 00A 000000A (No. AAAAAAA-xxxxx) and Blind-000000B 00000B 000B 00B 000000B (No. BBBBBBB-xxxxx).}

%% file: section/97reference.tex
%% produce the bibliography for the citations in your paper.
% https://www.overleaf.com/learn/latex/Bibtex_bibliography_styles#Table_of_stylename_values

\onecolumn
\begin{multicols}{2}

\bibliographystyle{abbrv} % or {acm}
\bibliography{section/97ref}  % *.bib

\end{multicols}

%% file: main.bbl
\begin{thebibliography}{10}

\bibitem{dram-exchange}
{A Business Division of TrendForce Corp}.
\newblock {DRAMeXchange}.
\newblock \burl{https://www.dramexchange.com}, 2022 (accessed 18 January 2023).

\bibitem{rtas16alhammad}
A.~Alhammad and R.~Pellizzoni.
\newblock {Trading Cores for Memory Bandwidth in Real-Time Systems}.
\newblock In {\em Proceedings of the Real-Time and Embedded Technology and
  Applications Symposium (RTAS'16)}, pages 1--11, Vienna, Austria, 2016. IEEE.

\bibitem{bankmycell}
BankMycell.
\newblock {How Many Smartphones are in the World?}
\newblock
  \burl{https://www.bankmycell.com/blog/how-many-phones-are-in-the-world}, 2022
  (accessed 18 January 2023).

\bibitem{issc13-blockio}
M.~Bj{\o}rling, J.~Axboe, D.~Nellans, and P.~Bonnet.
\newblock {Linux Block IO: Introducing Multi-Queue SSD Access on Multi-core
  Systems}.
\newblock In {\em Proceedings of the International Systems and Storage
  Conference (SYSTOR'13)}, pages 1--10, Haifa, Israel, 2013. USENIX.

\bibitem{rtas20casini}
D.~Casini, A.~Biondi, G.~Nelissen, and G.~Buttazzo.
\newblock {A Holistic Memory Contention Analysis for Parallel Real-Time Tasks
  Under Partitioned Scheduling}.
\newblock In {\em Proceedings of the Real-Time and Embedded Technology and
  Applications Symposium (RTAS'20)}, pages 239--252, Sydney, Australia, 2020.
  IEEE.

\bibitem{ram-plus-ac}
A.~Central.
\newblock {Samsung Galaxy S21 phones are getting a memory boost you barely
  need}.
\newblock
  \burl{https://www.androidcentral.com/samsung-galaxy-s21-ram-plus-update},
  2022 (accessed 18 January 2023).

\bibitem{atc99-swap-compression}
R.~Cervera, T.~Cortes, and Y.~Becerra.
\newblock {Improving Application Performance through Swap Compression}.
\newblock In {\em Proceedings of the Annual Technical Conference (ATC'99)},
  pages 1--12, Monterey, CA, USA, 1999. USENIX.

\bibitem{emsoft16-swap-dram-nvm}
X.~Chen, E.~H.-M. Sha, W.~Jiang, Q.~Zhuge, J.~Chen, J.~Qin, and Y.~Zeng.
\newblock {The Design of an Efficient Swap Mechanism for Hybrid DRAM-NVM
  Systems}.
\newblock In {\em Proceedings of the International Conference on Embedded
  Software (EMSOFT'16)}, pages 1--10, Pittsburgh, PA, USA, 2016. ACM.

\bibitem{lz4}
L.~Community.
\newblock {LZ4 - Extremely Fast Compression}.
\newblock \burl{https://github.com/lz4/lz4}, 2022 (accessed 18 January 2023).

\bibitem{lwn10-another-oomk}
J.~Corbet.
\newblock {Another OOM Killer Rewrite}.
\newblock \burl{https://lwn.net/Articles/391222/}, 2010 (accessed 18 January
  2023).

\bibitem{linux-zswap}
J.~Corbet.
\newblock {Zswap: Compressed SWAP Caching}.
\newblock \burl{https://lwn.net/Articles/528817/}, 2012 (accessed 18 January
  2023).

\bibitem{swapping}
J.~Corbet.
\newblock {Reconsidering Swapping}.
\newblock \burl{https://lwn.net/Articles/690079/}, 2016 (accessed 18 January
  2023).

\bibitem{lwn12-us-lmkd}
J.~Corbet.
\newblock {User-Space Low-Memory Killer Daemon}.
\newblock \burl{https://lwn.net/Articles/511731/}, 2016 (accessed 18 January
  2023).

\bibitem{mobicom17-ad-measure}
M.~D. Corner, B.~N. Levine, O.~Ismail, and A.~Upreti.
\newblock {Advertising-Based Measurement: A Platform of 7 Billion Mobile
  Devices}.
\newblock In {\em Proceedings of the International Conference on Mobile
  Computing and Networking (Mobicom'17)}, pages 435--447, Snowbird, Utah, USA,
  2017. ACM.

\bibitem{microcom16-zram-dedup}
S.~Desireddy and D.~R. Pathireddy.
\newblock {Optimize In-Kernel Swap Memory by Avoiding Duplicate Swap-out
  Pages}.
\newblock In {\em Proceedings of the International Conference on
  Microelectronics, Computing and Communications (MicroCom'16)}, pages 1--4,
  Durgapur, India, 2016. IEEE.

\bibitem{mobicom15-caredroid}
S.~Elmalaki, L.~Wanner, and M.~Srivastava.
\newblock {CAreDroid: Adaptation Framework for Android Context-Aware
  Applications}.
\newblock In {\em Proceedings of the International Conference on Mobile
  Computing and Networking (Mobicom'15)}, pages 386--399, Paris, France, 2015.
  ACM.

\bibitem{android-aosp}
Google.
\newblock {Android Open-Source Project (AOSP)}.
\newblock \burl{https://source.android.com}, 2008 (accessed 18 January 2023).

\bibitem{android-dev}
Google.
\newblock {Android Developers}.
\newblock \burl{https://developer.android.com}, 2012 (accessed 18 January
  2023).

\bibitem{android-activity-manager}
Google.
\newblock {Android Developers: Activity Manager}.
\newblock
  \burl{https://developer.android.com/reference/android/app/ActivityManager},
  2012 (accessed 18 January 2023).

\bibitem{android-aapt}
Google.
\newblock {Android Developers: AAPT2 (Android Asset Packaging Tool)}.
\newblock \burl{https://developer.android.com/studio/command-line/aapt2}, 2013
  (accessed 18 January 2023).

\bibitem{android-ndk}
Google.
\newblock {Android Native Development Kit (NDK)}.
\newblock \burl{https://developer.android.com/ndk/}, 2015 (accessed 18 January
  2023).

\bibitem{android-mem-alloc}
Google.
\newblock {Memory Allocation among Processes}.
\newblock
  \burl{https://developer.android.com/topic/performance/memory-management},
  2016 (accessed 18 January 2023).

\bibitem{android-app-startup-time}
Google.
\newblock {Android Developers: App Startup Time}.
\newblock
  \burl{https://developer.android.com/topic/performance/vitals/launch-time},
  2018 (accessed 18 January 2023).

\bibitem{android-hierarchy-viewer}
Google.
\newblock {Android Developers: Profile Your Layout with Hierarchy Viewer}.
\newblock \burl{https://developer.android.com/studio/profile/hierarchy-viewer},
  2018 (accessed 18 January 2023).

\bibitem{android-lmkd}
Google.
\newblock {Android Low-Memory Killer Daemon (LMKD)}.
\newblock \burl{https://source.android.com/devices/tech/perf/lmkd}, 2019
  (accessed 18 January 2023).

\bibitem{google-play}
Google.
\newblock {Android Apps on Google Play}.
\newblock \burl{https://play.google.com/store/apps/}, 2022 (accessed 18 January
  2023).

\bibitem{android-noti}
Google.
\newblock {Android Developers: Notifications}.
\newblock
  \burl{https://developer.android.com/guide/topics/ui/notifiers/notifications},
  2022 (accessed 18 January 2023).

\bibitem{android-services}
Google.
\newblock {Android Developers: Services}.
\newblock \burl{https://developer.android.com/guide/components/services}, 2022
  (accessed 18 January 2023).

\bibitem{app-hiber}
Google.
\newblock {App Hibernation}.
\newblock \burl{https://source.android.com/devices/tech/perf/hiber}, 2022
  (accessed 18 January 2023).

\bibitem{android-factory-image}
Google.
\newblock {Factory Images for Nexus and Pixel Devices}.
\newblock \burl{https://developers.google.com/android/images}, 2022 (accessed
  18 January 2023).

\bibitem{google-pixel6}
Google.
\newblock {Google Pixel6}.
\newblock \burl{https://store.google.com/category/phones}, 2022 (accessed 18
  January 2023).

\bibitem{atc18-fasttrack}
S.~S. Hahn, S.~Lee, I.~Yee, D.~Ryu, and J.~Kim.
\newblock {Fasttrack: Foreground App-aware I/O Management for Improving User
  Experience of Android Smartphones}.
\newblock In {\em Proceedings of the Annual Technical Conference (ATC'18)},
  pages 15--28, Boston, MA, USA, 2018. USENIX.

\bibitem{access18-hybrid-swap}
J.~Han, S.~Kim, S.~Lee, J.~Lee, and S.~J. Kim.
\newblock {A Hybrid Swapping Scheme based on Per-Process Reclaim for
  Performance Improvement of Android Smartphones}.
\newblock {\em IEEE Access}, 6:56099--56108, 2018.

\bibitem{atc16-linux-mm-evolution}
J.~Huang, M.~K. Qureshi, and K.~Schwan.
\newblock {An Evolutionary Study of Linux Memory Management for Fun and
  Profit}.
\newblock In {\em Proceedings of the Annual Technical Conference (ATC'16)},
  pages 465--478, Denver, CO, USA, 2016. USENIX.

\bibitem{opensource-huawei}
Huawei.
\newblock {Huawei Open-Source Release Center}.
\newblock \burl{https://consumer.huawei.com/en/opensource/}, 2013 (accessed 18
  January 2023).

\bibitem{android-market-share-idc}
IDC.
\newblock {Smartphone Market Share}.
\newblock \burl{https://www.idc.com/promo/smartphone-market-share}, 2022
  (accessed 18 January 2023).

\bibitem{mobile-computing-09cont}
S.~Ioannidis, A.~Chaintreau, and L.~Massouli{\'e}.
\newblock {Distributing Content Updates over a Mobile Social Network}.
\newblock {\em ACM SIGMOBILE Mob. Comput. Commun. Rev.}, 13:44--47, 2009.

\bibitem{simform21-private}
M.~Kataria.
\newblock {App Usage Statistics 2021 that will Surprise You (Private webpage)}.
\newblock \burl{https://www.simform.com/contact/}, 2022 (accessed 18 January
  2023).

\bibitem{simform21-public}
M.~Kataria.
\newblock {App Usage Statistics 2021 that will Surprise You (Public webpage)}.
\newblock \burl{https://www.simform.com/blog/the-state-of-mobile-app-usage/},
  2022 (accessed 18 January 2023).

\bibitem{access20-selective-swap-kill}
J.~Kim and H.~Bahn.
\newblock {Maintaining Application Context of Smartphones by Selectively
  Supporting Swap and Kill}.
\newblock {\em IEEE Access}, 8:85140--85153, 2020.

\bibitem{access19-ezswap}
J.~Kim, C.~Kim, and E.~Seo.
\newblock {Ezswap: Enhanced Compressed Swap Scheme for Mobile Devices}.
\newblock {\em IEEE Access}, 7:139678--139691, 2019.

\bibitem{tecs16-smart-lmk}
S.-H. Kim, J.~Jeong, J.-S. Kim, and S.~Maeng.
\newblock {SmartLMK: A Memory Reclamation Scheme for Improving User-Perceived
  App Launch Time}.
\newblock {\em ACM Trans. Embed. Comput. Syst. (TECS)}, 15(3):1--25, 2016.

\bibitem{mobicom18android}
N.~Klugman, V.~Jacome, M.~Clark, M.~Podolsky, P.~Pannuto, N.~Jackson, A.~S.
  Nassor, C.~Wolfram, D.~Callaway, J.~Taneja, and P.~Dutta.
\newblock {Experience: Android Resists Liberation from Its Primary Use Case}.
\newblock In {\em Proceedings of the International Conference on Mobile
  Computing and Networking (Mobicom'18)}, pages 540--556, New Delhi, India,
  2018. ACM.

\bibitem{mobicom15chall}
J.~F. Kurose.
\newblock {Research Challenges and Opportunities in a Mobility-centric World}.
\newblock In {\em Proceedings of the International Conference on Mobile
  Computing and Networking (Mobicom'15)}, page 290, Paris, France, 2015. ACM.

\bibitem{tce10-swap-nand}
O.~Kwon and K.~Koh.
\newblock {Swap Space Management Technique for Portable Consumer Electronics
  with NAND Flash Memory}.
\newblock {\em IEEE Trans. Consum. Electron. (TCE)}, 56(3):1524--1531, 2010.

\bibitem{atc20-marvin}
N.~Lebeck, A.~Krishnamurthy, H.~M. Levy, and I.~Zhang.
\newblock {End the Senseless Killing: Improving Memory Management for Mobile
  Operating Systems}.
\newblock In {\em Proceedings of the Annual Technical Conference (ATC'20)},
  pages 873--887. USENIX, 2020.

\bibitem{linker-loader}
J.~R. Levine.
\newblock {\em {Linkers \& Loaders}}.
\newblock Morgan Kaufmann, 1999.

\bibitem{iwcmc17-opt-lmk-ml}
C.~Li, J.~Bao, and H.~Wang.
\newblock {Optimizing Low Memory Killers for Mobile Devices using Reinforcement
  Learning}.
\newblock In {\em Proceedings of International Wireless Communications and
  Mobile Computing Conference (IWCMC'17)}, pages 2169--2174, Valencia, Spain,
  2017. IEEE.

\bibitem{tcaics20-seal}
C.~Li, L.~Shi, Y.~Liang, and C.~J. Xue.
\newblock {SEAL: User Experience-Aware Two-Level Swap for Mobile Devices}.
\newblock {\em IEEE Trans. Comput.-Aided Des. Integr. Circuits Syst.},
  39(11):4102--4114, 2020.

\bibitem{tc20-zweilous}
G.~Li, W.~Chen, and Y.~Xiang.
\newblock {Zweilous: A Decoupled and Flexible Memory Management Framework}.
\newblock {\em ACM Trans. Comput. (TC)}, 2020.

\bibitem{swam}
G.~Lim.
\newblock {SWAM}.
\newblock \burl{https://mobile-swam.github.io/}, 2023 (accessed 18 January
  2023).

\bibitem{tce13-vnode}
G.~Lim, C.~Min, and Y.~I. Eom.
\newblock {Virtual Memory Partitioning for Enhancing Application Performance in
  Mobile Platforms}.
\newblock {\em IEEE Trans. Consum. Electron. (TCE)}, 59(4):786--794, 2013.

\bibitem{tce12-flash-swap}
M.~Lin and S.~Chen.
\newblock {Flash-Aware Linux Swap System for Portable Consumer Electronics}.
\newblock {\em IEEE Trans. Consum. Electron. (TCE)}, 58(2):419--427, 2012.

\bibitem{rtss17nasri}
M.~Nasri and B.~B. Brandenburg.
\newblock {An Exact and Sustainable Analysis of Non-Preemptive Scheduling}.
\newblock In {\em Proceedings of the Real-Time Systems Symposium (RTSS'17)},
  pages 12--23, Paris, France, 2017. IEEE.

\bibitem{mobisys15-app-rw-isolation}
D.~T. Nguyen, G.~Zhou, G.~Xing, X.~Qi, Z.~Hao, G.~Peng, and Q.~Yang.
\newblock {Reducing Smartphone Application Delay Through Read/Write Isolation}.
\newblock In {\em Proceedings of the Annual International Conference on Mobile
  Systems, Applications, and Services (MobiSys'15)}, pages 287--300, Florence
  Italy, 2015. ACM.

\bibitem{opensource-oppo}
Oppo.
\newblock {Oppo Open-Source Release Center}.
\newblock \burl{https://github.com/oppo-source/}, 2014 (accessed 18 January
  2023).

\bibitem{ram-expansion-oppo}
Oppo.
\newblock {Introducing RAM Expansion Feature (Oppo Australia)}.
\newblock \burl{https://support.oppo.com/au/answer/?aid=2033453}, 2019
  (accessed 18 January 2023).

\bibitem{ram-plus}
SamMobile.
\newblock {RAM Plus: Samsung’s Extra RAM Feature is Arriving on More Phones,
  both Mid-range and Flagship}.
\newblock
  \burl{https://www.sammobile.com/news/samsung-galaxy-a52s-5g-virtual-plus-feature/},
  2022 (accessed 18 January 2023).

\bibitem{opensource-samsung}
Samsung.
\newblock {Samsung Open-Source Release Center}.
\newblock \burl{https://opensource.samsung.com/}, 2012 (accessed 18 January
  2023).

\bibitem{galaxy-a12}
Samsung.
\newblock {Galaxy A12}.
\newblock
  \burl{https://www.samsung.com/africa_en/smartphones/galaxy-a/galaxy-a12-black-64gb-sm-a125fzkgxfe/},
  2022 (accessed 18 January 2023).

\bibitem{eufs-samsung}
Samsung.
\newblock {Samsung Begins Mass Production of the Fastest Storage for Flagship
  Smartphones}.
\newblock
  \burl{https://news.samsung.com/global/samsung-begins-mass-production-of-the-fastest-storage-for-flagship-smartphones},
  2022 (accessed 18 January 2023).

\bibitem{galaxy-s}
Samsung.
\newblock {Samsung Galaxy S}.
\newblock \burl{https://www.samsung.com/us/mobile/phones/galaxy-s/}, 2023
  (accessed 18 January 2023).

\bibitem{atc10-flashvm}
M.~Saxena and M.~M. Swift.
\newblock {FlashVM: Virtual Memory Management on Flash}.
\newblock In {\em Proceedings of the Annual Technical Conference (ATC'10)},
  pages 1--14, Boston, MA, USA, 2010. USENIX.

\bibitem{atc21-asap}
S.~Son, S.~Y. Lee, Y.~Jin, J.~Bae, J.~Jeong, T.~J. Ham, J.~W. Lee, and H.~Yoon.
\newblock {ASAP: Fast Mobile Application Switch via Adaptive Prepaging}.
\newblock In {\em Proceedings of the Annual Technical Conference (ATC'21)},
  pages 365--380, 2021.

\bibitem{tecs14-person-opt}
W.~Song, Y.~Kim, H.~Kim, J.~Lim, and J.~Kim.
\newblock {Personalized Optimization for Android Smartphones}.
\newblock {\em ACM Trans. Embed. Comput. Syst.}, 13(2s):1--25, 2014.

\bibitem{linux-zram}
L.~Torvalds.
\newblock {ZRAM}.
\newblock \burl{https://www.kernel.org/doc/Documentation/blockdev/zram.txt},
  2022 (accessed 18 January 2023).

\bibitem{opensource-xiaomi}
XiaoMi.
\newblock {XiaoMi Open-Source Release Center}.
\newblock \burl{https://xiaomi.github.io/}, 2014 (accessed 18 January 2023).

\bibitem{mobisys12-fast-app-launching}
T.~Yan, D.~Chu, D.~Ganesan, A.~Kansal, and J.~Liu.
\newblock {Fast App Launching for Mobile Devices Using Predictive User
  Context}.
\newblock In {\em Proceedings of the International Conference on Mobile
  Systems, Applications, and Services (MobiSys'12)}, pages 113--126, New York,
  NY, USA, 2012. ACM.

\bibitem{emsoft14-nvram}
K.~Zhong, T.~Wang, X.~Zhu, L.~Long, D.~Liu, W.~Liu, Z.~Shao, and E.~H.-M. Sha.
\newblock {Building High-performance Smartphones via Non-volatile Memory: The
  Swap Approach}.
\newblock In {\em Proceedings of the International Conference on Embedded
  Software (EMSOFT'14)}, pages 1--10, Uttar Pradesh, India, 2014. ACM.

\bibitem{mobicom14use}
P.~Zhou, M.~Li, and G.~Shen.
\newblock {Use It Free: Instantly Knowing Your Phone Attitude}.
\newblock In {\em Proceedings of the International Conference on Mobile
  Computing and Networking (Mobicom'14)}, pages 605--616, Maui, Hawaii, USA,
  2014. ACM.

\bibitem{dac17-smartswap}
X.~Zhu, D.~Liu, K.~Zhong, J.~Ren, and T.~Li.
\newblock {Smartswap: High-Performance and User Experience Friendly Swapping in
  Mobile Systems}.
\newblock In {\em Proceedings of the Annual Design Automation Conference
  (DAC'17)}, pages 1--6, Austin, TX, USA, 2017. ACM.

\end{thebibliography}
